\documentclass[a4paper,fleqn,usenatbib]{mnras}

\usepackage[dvipsnames]{xcolor}

\usepackage[T1]{fontenc}
\usepackage{ae,aecompl}

\usepackage{graphicx}	
\usepackage{amsmath}	
\usepackage{amssymb}	
\usepackage{newtxtext,newtxmath}

\title[Convection in magnetically arrested discs]{What really makes an accretion disc MAD}

\author[M. Begelman et al.]{
Mitchell C. Begelman,$^{1,2}$\thanks{E-mail: mitch@jila.colorado.edu}
Nicolas Scepi,$^{1}$
and Jason Dexter$^{1,2}$
\\
$^{1}$JILA, University of Colorado and National Institute of Standards and Technology, 440 UCB, Boulder, CO 80309-0440, USA \\
$^{2}$Department of Astrophysical and Planetary Sciences, 391 UCB, Boulder, CO 80309-0391, USA
}

\date{Accepted XXX. Received YYY; in original form ZZZ}

\pubyear{2020}

\begin{document}
\label{firstpage}
\pagerange{\pageref{firstpage}--\pageref{lastpage}}
\maketitle

\begin{abstract}
Magnetically arrested accretion discs (MADs) around black holes (BH) have the potential to stimulate the production of powerful jets and account for recent ultra-high-resolution observations of BH environments.  Their main properties are usually attributed to the accumulation of dynamically significant net magnetic (vertical) flux throughout the arrested region, which is then regulated by interchange instabilities.  Here we propose instead that it is mainly a dynamically important {\it toroidal} field --- the result of dynamo action triggered by the significant but still relatively weak vertical field --- that defines and regulates the properties of MADs. We suggest that rapid convection-like instabilities, involving interchange  of toroidal flux tubes and  operating concurrently with the magnetorotatonal instability (MRI), can regulate the structure of the disc and the escape of net flux. We generalize the convective stability criteria and disc structure equations to include the effects of a strong toroidal field and show that convective flows could be driven towards two distinct marginally stable states, one of which we associate with MADs.  We confirm the plausibility of our theoretical model by comparing its quantitative predictions to simulations of both MAD and SANE (strongly magnetized but not ``arrested'') discs, and suggest a set of criteria that could help to distinguish MADs from other accretion states.  Contrary to previous claims in the literature, we argue that MRI is not suppressed in MADs and is probably responsible for the existence of the strong toroidal field.

\end{abstract}




\section{Introduction}

Angular momentum transport in black hole (BH) accretion discs is likely governed by the magnetorotational instability (MRI) \citep{balbus1991,balbus1998}. The strengths of the resulting saturated magnetic field and Maxwell stress increase with increasing accreted net magnetic flux \citep{hawley1995,bai2013,salvessen2016,scepi2018b,mishra2020}. In addition, MHD simulations find qualitatively different behavior when a large amount of magnetic flux is accreted: the flux threading the central black hole saturates, with magnetically-dominated plasma expelled out to larger radius \citep{igumenshchev2008,tchekhovskoy2011,porth2021}. The inner accretion flow is compressed by a wide, dynamically important magnetosphere \citep{mckinney2012}, and the \citet{blandford1977} process can extract black hole spin energy in the form of powerful jets. The flow is also substantially sub-Keplerian, and shows non-axisymmetric structure in the mass density, magnetic field strength, and gas temperature \citep{igumenshchev2008,mckinney2012,porth2021}.

While the empirical properties of these magnetically arrested discs (MADs) are well-established from simulations, the physical mechanisms causing magnetic flux saturation and determining their steady state structure remain unclear. By analogy with accretion onto strongly magnetized neutron stars, \citet{narayan2003} proposed that MAD accretion may be truncated at a magnetospheric radius where the ram pressure of accreting material is balanced by the pressure of the poloidal magnetic field. In axisymmetry such a standoff may be stable, but in three dimensions accretion could then proceed via non-axisymmetric Rayleigh-Taylor (or ``interchange'') instabilities \citep{igumenshchev2008,mckinney2012,avara2016,marshall2018} where the gas slips past the strong vertical magnetic field. In the limit of such strong vertical fields, it is unclear what role the MRI plays in transporting angular momentum \citep{mckinney2012,marshall2018,white2019}.

A major source of tension between these theoretical ideas and simulations is the relative insignificance of the net vertical field strength seen in the latter, compared to the strong, organized toroidal field ($B_\phi$) that dominates the time- and azimuthally averaged magnetic structure from very close to the black hole to the largest radii modeled.  It is hard to see how the vertical field can drive interchange instabilities when the actual field structure is overwhelmingly dominated by azimuthal hoops of field and the net field structure is helical with a small pitch angle.  On the other hand, it has been argued that an instability involving the vertical field is necessary to drive the strong turbulence that is observed, given the possible suppression of MRI due to the strength of the vertical field \citep{mckinney2012,marshall2018,white2019} and the pinching of the inner flow by the pressure of the jet \citep{mckinney2012}.

In this paper, we argue that the structure of MADs and the saturation of their net magnetic flux are indeed regulated by interchange-type instabilities, but not the kind that are usually assumed.  We show that, under the conditions found in simulations of MADs, strong convective instabilities --- inducing primarily radial motions --- can operate on dynamical timescales.  The instabilities are basically axisymmetric and involve the interchange of {\it toroidal} flux tubes. They are driven by the combined effects of buoyancy due to gas pressure gradients and the strong toroidal magnetic field, and have nothing directly to do with the much weaker vertical field. However, the vertical field is presumably necessary to stimulate the creation of the organized toroidal field through dynamo action.  Thus, the vertical field plays a dual, though indirect, role: 1) it must be sufficiently strong to catalyze the development of a convectively unstable toroidal field and 2) once strong convection is established, it behaves as a passive contaminant that can be transported convectively as the flow evolves toward marginal stability. In the following we will refer to the instability as convection although it also involves stresses associated with the toroidal field.

The plan of the paper is as follows. In \autoref{sec:sims} we analyze data from a long-duration and large-dynamic-range, 3D, global, general relativistic MHD (GRMHD) simulation of MAD accretion to show that the vertical magnetic field is much weaker than the azimuthal field at all radii. The pressure of the vertical field is not sufficient to balance the ram pressure of infalling material, and interchange instabilities based on the poloidal field therefore seem unlikely to operate. Evidently, some other mechanism is needed to explain the flux saturation and steady-state structure of MADs.   We next analyze the convective instability of rotating flows containing a dynamically significant toroidal magnetic field, generalizing the two well-known H\o iland criteria (\autoref{sec:Hoiland}). In \autoref{sec:selfsim} we derive self-similar models of strongly magnetized accretion flows that are marginally unstable to radial or vertical convection. The models that are marginally unstable to radial convection, in particular, provide a number of quantitative  predictions that closely match our long duration MAD GRMHD model (\autoref{sec:MADsim}), but not a long duration SANE (Standard and Normal Evolution) model that lacks a saturated field.  Our results thus suggest that convective instabilities may play an important role in determining the structure and dynamics of MADs (\autoref{sec:discussion}).

Our interpretation of MAD structure requires a dynamically important toroidal magnetic field, but does not specify how that field arises in the first place.  MRI provides an attractive mechanism to create a large-scale toroidal field through dynamo action.  To address concerns that MRI may be suppressed in MADs \citep{mckinney2012,white2019}, in \autoref{sec:MRI} we consider a general dispersion relation for MRI \citep{das2018} under MAD conditions and show that MADs likely remain unstable to a range of MRI modes at all radii, especially when modes with non-zero azimuthal wavenumber are taken into account.  

\section{Properties of magnetically arrested discs from GRMHD simulations}
\label{sec:sims}

In this section, we summarize the empirical knowledge on MAD simulations as well as present new analysis from a long duration MAD simulation that shows in which aspects previous models of MADs fail to correctly describe them.

\subsection{Long-duration GRMHD simulations}
\label{sec:sims_setup}

To analyze our simulations up to very large radii, we ran a MAD and a SANE simulation up to 90,000 $r_g/c$ and 190,000 $r_g/c$, respectively, so that they achieved inflow equilibrium up to $\approx100\:r_g$. Details of the two simulations can be found in \cite{dexter2020a,dexter2020b}.

The MAD and SANE simulations are 3D GRMHD simulations run with the public code \textsc{harmpi}\footnote{https://github.com/atchekho/harmpi} \citep{tchekhovskoy2019} that includes a scheme to evolve the electron entropy separately from the total (proton+electron) fluid \citep{ressler2015}. We use an adiabatic equation of state with a relativistic adiabatic index of $4/3$ for the electrons and a non-relativistic adiabatic index of $5/3$ for the proton+electron fluid. The MAD and SANE simulations were initialized from a Fishbone-Moncrief torus \citep{fishbone1976} with an inner radius of 12 $r_g$ and a pressure maximum at 25 $r_g$.  The electron internal energy density is initially $0.1$ that of the fluid.

The grids are based on spherical-polar Kerr-Schild coordinates, which are stretched in the $\theta$-direction. The grid resolution is $320\times256\times160$ in the $r,\theta$ and $\phi$ directions, respectively. The magnetic field configuration is initialized as a single poloidal field loop, whose radial profile is chosen to produce either a MAD or SANE accretion flow. The MAD (SANE) simulation used a black hole spin parameter of $a=0.9375$ ($a=0$). 

\subsection{Time-dependent and non-axisymmetric behavior of MADs: the role of the eruptive bubbles}\label{sec:time_dep}

We define the time-averaged spherical magnetic flux as 
\begin{equation}
    \Phi_\mathrm{sph}(r) = \int_{\mathcal{H}(r)}\sqrt{4\pi}B^r\sqrt{-g}d\theta d\phi,
\end{equation}
where $\mathcal{H}(r)$ is a hemisphere of radius $r$ and $B^r$ is the radial component of the 3-vector magnetic field. We also define the time-averaged midplane magnetic flux as 
\begin{equation}
    \Phi_\mathrm{mid}(r) = \int\sqrt{4\pi}B^\theta(\theta=\pi/2)\sqrt{-g}dr d\phi,
\end{equation}
where $B^\theta$ is the latitudinal component of the 3-vector magnetic field. Time averages are always made between $70,000\:r_g/c$ and $90,000\:r_g/c$ for our MAD simulation and between $170,000\:r_g/c$ and $190,000\:r_g/c$ for our SANE simulation, unless otherwise specified. 

\begin{figure}
\includegraphics[width=85mm]{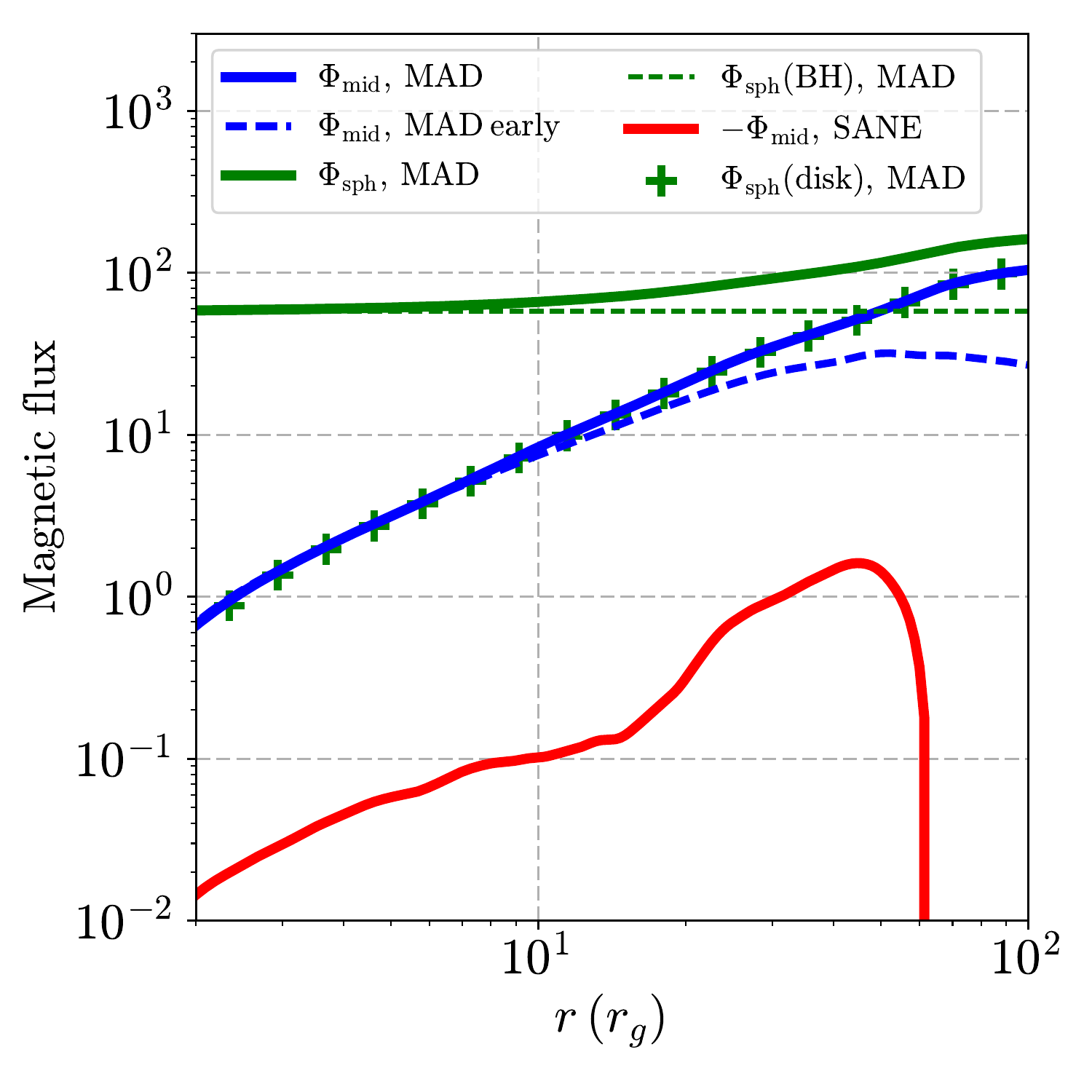}
\vspace{-24pt}
\caption{Midplane magnetic flux for our MAD simulation (blue solid line) and SANE simulation (red solid line) as a function of radius. The dashed blue line shows the midplane magnetic flux for our MAD simulation at an earlier time, between $20,000$ and $30,000$ $r_g/c$. We also show the spherical magnetic flux for our MAD simulation (green solid line) as a function of radius, with the green dashed line and chain of plus-signs showing the contributions of fluxes threading the BH and the disc, respectively. When we subtract the contribution from the BH, the spherical magnetic flux in the disc and the midplane magnetic flux are identical (as expected) and scale $\propto r$. A comparison of the solid and dashed blue lines shows that the magnetic flux within a given disc radius accumulates with time until it saturates.}
\label{fig:flux_steady}
\end{figure}

Because of the conservation of magnetic flux, $\Phi_\mathrm{sph}(r)$ should be exactly equal to $\Phi_\mathrm{mid}(r)$ plus the flux threading the black hole. This is exactly what we find in  \autoref{fig:flux_steady}, where the green line represents the time-averaged total (disc+BH) spherical magnetic flux for our MAD simuation and the blue line represents the time-averaged midplane magnetic flux. The green chain of plus-signs shows the disc contribution to the spherical magnetic flux and is exactly equal to the midplane magnetic flux.

\begin{figure*}
\includegraphics[width=\textwidth]{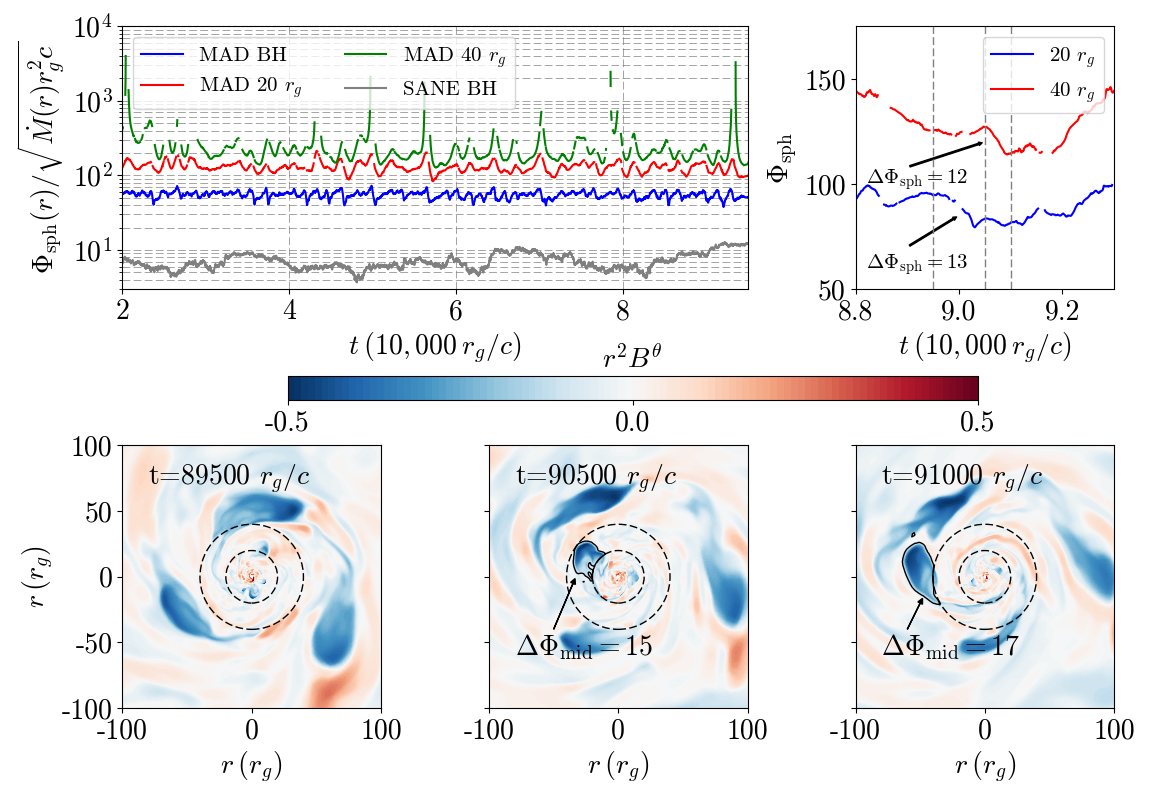}
\caption{Top left panel: Dimensionless magnetic flux as a function of time. The blue, red and green curves show the enclosed flux in the MAD simulation at the BH, at a radius of 20 $r_g$ and at a radius of 40 $r_g$, respectively. In MADs the flux saturates at the BH as well as in the disc at larger radii. The grey solid curve shows the flux at the BH in the SANE simulation. Top right panel: Dimensionalized magnetic flux as a function of time zoomed between 88,000 $r_g/c$ and 92,500 $r_g/c$. The blue and red curves show the flux at 20 $r_g$ and 40 $r_g$, respectively. The vertical grey dashed lines show the times at which the snapshots of the bubbles (in the bottom panels) are made. We also indicate the drops in the flux, $\Delta \Phi_\mathrm{sph}$ between 89,500 $r_g/c$ and 90,500 $r_g/c$ and between 90,500 $r_g/c$ and 91,000 $r_g/c$, for comparison with the amount of flux in the bubbles. Bottom panels: Midplane cuts of $r^2 B^\theta$ at 89,500 $r_g/c$, 90,500 $r_g/c$ and 91,000 $r_g/c$ following the propagation of a magnetized, low-density bubble in the disc, highlighted by a black contour, as it crosses the 20 $r_g$ and 40 $r_g$ surfaces indicated by black dashed circles. We also indicate the amount of flux in the bubble, $\Delta \Phi_\mathrm{mid}$. $\Delta \Phi_\mathrm{mid}$ is very close to $\Delta \Phi_\mathrm{sph}$, suggesting that bubbles might account for most of the fluctuations in the magnetic flux at every radius in the disc.}
\label{fig:flux_saturation_BH}
\end{figure*}

A well-known feature of MADs is the saturation of the dimensionless magnetic flux threading the black hole, $\phi_\mathrm{BH} = \Phi_{\rm sph}\mathrm{(BH)}/\sqrt{\dot{M}_\mathrm{BH} r_g^2 c}$ \citep{tchekhovskoy2011}, at $\phi_\mathrm{BH} \simeq 50$ as can be seen in \autoref{fig:flux_steady} and the top left panel of \autoref{fig:flux_saturation_BH}. The dimensionless flux on the BH in the SANE simulation, also shown in the top left panel of \autoref{fig:flux_saturation_BH}, is much lower during the same period. It is generally accepted that the mechanism regulating the saturation of $\phi_\mathrm{BH}$ in MADs involves the formation of highly magnetized, low-density bubbles near the black hole that rise buoyantly to larger radii, taking the flux away \citep{igumenshchev2008,tchekhovskoy2011,mckinney2012,avara2016,marshall2018,porth2021}. As they rise the bubbles create an empty space near the black hole where the open magnetic field lines threading the black hole, which are associated with the jet, can reconnect \citep{igumenshchev2003,scepi2021,ripperda2021}. This reconnection effectively regulates the amount of magnetic flux threading the black hole.

It is less clear, however, how these bubbles participate in the regulation of the magnetic flux in the rest of the disc as they escape to larger radii. In the top left panel of \autoref{fig:flux_saturation_BH}, we show the dimensionless magnetic flux at $20$ and $40$ $r_g$. We see that, similarly to the magnetic flux on the black hole, the enclosed magnetic flux in the disc saturates at a constant, radius-dependent value around which it oscillates. This can also be seen in \autoref{fig:flux_steady} where a comparison of the dashed and solid blue lines shows that the magnetic flux in the disc saturates first in the inner parts of the disc and then gradually builds up in the outer disc to reach the saturated value.

In \autoref{fig:flux_saturation_BH} we investigate the contribution of the magnetized, empty bubbles created close to the BH to the saturation of the flux in the disc as a function of radius. We follow one bubble, highlighted by a black solid line in the bottom panels, as it propagates outward by plotting $r^2 B^\theta$ in the midplane, We indicate the surfaces $r=20$ and $40$ $r_g$ as black dashed circles and isolate two moments in time when the bubble crosses these surfaces. We estimate the amount of magnetic flux in the bubble as
\begin{equation}
    \Delta \Phi_\mathrm{mid} = \int_{\mathrm{bubble}}\sqrt{4\pi}B^\theta(\theta=\pi/2)\sqrt{-g}dr d\phi.
\end{equation}
We find that the magnetic flux carried by this bubble does not change significantly as it moves outward, since it carries a magnetic flux of $\approx 15$ in code units when located at $20$ $r_g$ and $\approx 17$ in code units when located at $40$ $r_g$. However, the surface area of the bubble increases, implying lower magnetic field intensities at larger radii \citep{porth2021}.

In the top right panel of  \autoref{fig:flux_saturation_BH} we show the evolution of $\Phi_\mathrm{sph}$ at $20$ and $40$ $r_g$ during the escape of the bubble. We see that the amount of flux carried in the bubble, $\Delta \Phi_\mathrm{mid}$, as it crosses the $20$ and $40$ $r_g$ surfaces, at $90,500$ and $91,000$ $r_g/c$, respectively, matches quite well the drop of $\Phi_\mathrm{sph}$ observed at the corresponding time and radius. Thus, we believe that the fluctuations in $\Phi_\mathrm{sph}$ at large radii can be accounted for entirely by the bubbles.

However, accounting for the fluctuations in the magnetic flux of MADs does not automatically imply that the escape of bubbles can explain the steady-state radial profile of the flux. In \autoref{fig:flux_steady} we see that the steady-state saturated value of $\Phi_\mathrm{mid}$ increases outward approximately $\propto r$. Given that the magnetic flux carried by a bubble does not increase as it moves outward, it is impossible for the bubbles to account for the steady-state radial profile of the magnetic flux. We believe that the role of the bubbles is simply to transport the excess flux from the BH and, when they rise buoyantly outward, to induce fluctuations in the flux at larger radii. Hence, another mechanism must be at work to explain the profile of the magnetic flux.

\subsection{Steady-state behavior of MADs: sub-Keplerian rotation, dominant toroidal field and the magnetospheric radius}
\label{sec:sims_measurements}

\begin{figure}
\includegraphics[width=85mm]{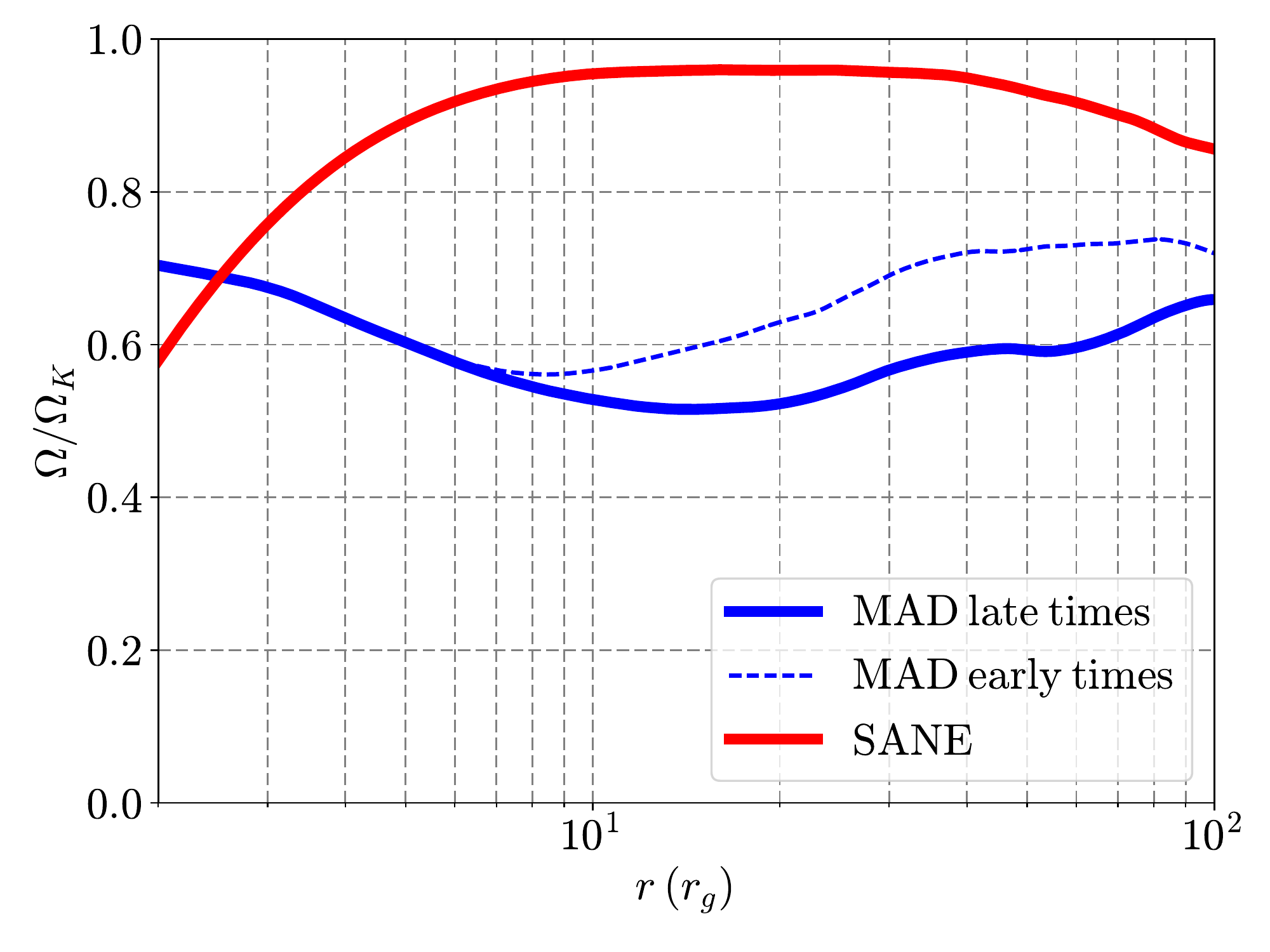}
\caption{Rotation rate weighted by density and averaged over time, $\theta$ and $\phi$, normalized to the Keplerian frequency, as a function of radius for the MAD (blue line) and SANE (red line) simulations.The dashed blue line shows the rotation profile from the MAD simulation averaged at an earlier time between $20,000$ and $30,000$ $r_g/c$. The MAD simulation is substantially sub-Keplerian while the SANE simulation is close to Keplerian. With time, the MAD becomes sub-Keplerian to larger and larger radii.}  
\label{fig:rotation}
\end{figure}

A strong characteristic of the steady-state structure of MADs is their substantially sub-Keplerian rotation. As can be seen in \autoref{fig:rotation}, the rotation profile in the MAD is built gradually. The disc first becomes sub-Keplerian in the inner parts and then in the outer parts as inflow equilibrium is established to larger radii.\footnote{At early times, the rotation rate at large radii in both MAD and SANE simulations is suppressed due to pressure support in the initial torus. The evolution of the angular velocity shown in \autoref{fig:rotation} would be even more pronounced if the simulation were initialized from a disc.} The sub-Keplerian rotation of MADs is what allows the magnetized bubbles studied in \autoref{sec:time_dep} to rise buoyantly and is often attributed to the poloidal magnetic field accumulated near the BH being so strong that it hinders the accretion flow \citep{narayan2003}. Indeed, \cite{narayan2003} and \cite{mckinney2012} define the magnetospheric radius, $r_\mathrm{mag}$, (the radius up to which the disc is MAD) as the radius where gravity is balanced by the poloidal magnetic tension so that
\begin{equation}
    \frac{G M\Sigma}{r_\mathrm{mag}^2} \sim \frac{2 B_r B_z}{4\pi} 
\end{equation}
where it is generally assumed that $B_r\sim B_z$. For the simulation studied here, we instead show below that the vertical magnetic field and poloidal magnetic tension are insufficient to balance the flow against gravity.
 
\begin{figure}
\includegraphics[width=85mm]{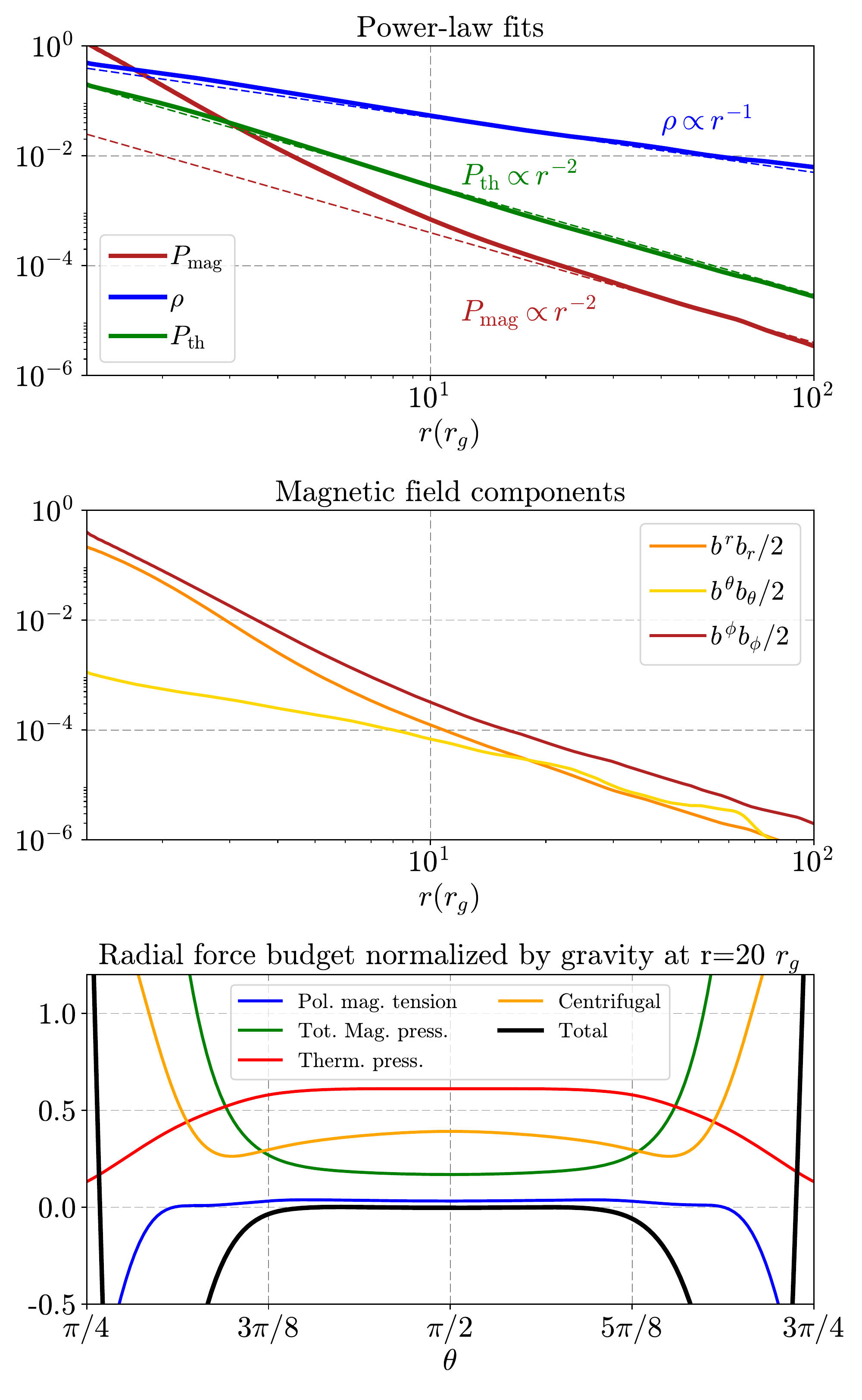}
\vspace*{-3mm}
\caption{Top panel: Radial dependence of quantities in our MAD simulation, weighted by density and averaged over time, $\theta$ and $\phi$. The density, thermal pressure and magnetic pressure are represented by the blue, green and red solid curves, respectively. The dashed blue, green and red curves indicate power-law fits for the density, thermal pressure and magnetic pressure. The fit for the magnetic pressure is made for $r>10\:r_g$. The power-law fits give an index $n=1/2$ according to equation (\ref{eq:selfsim}). Middle panel: Magnetic pressure due to the $r, \theta$ and $\phi$ components of the magnetic field, respectively as orange, yellow and red solid lines, as a function of radius. The toroidal field is the dominant component and the latitudinal field is sub-dominant. Bottom panel: Radial force budget normalized by gravity at a radius of 20 $r_g$ in our MAD simulation. The sum of all forces (including gravity, not shown), represented by the black solid line, should equal zero when the disc is in equilibrium. We find that this is true near the midplane of the disc. The blue, green, red and orange solid lines represent the poloidal magnetic tension, the gradient of magnetic pressure, the gradient of thermal pressure and the centrifugal force, respectively. We find that the centrifugal force and the gradient of thermal pressure are almost entirely balancing gravity.}
\label{fig:sim_forces}
\end{figure}

The top panel of \autoref{fig:sim_forces} shows radial profiles of the following density-weighted, spherically averaged quantities (denoted by $\langle \rangle_\rho$): density in blue, gas pressure in green, and magnetic pressure $b^\mu b_\mu$ in red, where $b^\mu$ is the 4-vector magnetic field. Gas pressure exceeds that of the magnetic field at all radii $r \gtrsim 2 \, r_g$. As can be seen from the middle panel of \autoref{fig:sim_forces} the field strength is dominated by the azimuthal contribution $b^\phi b_\phi$ at all radii, with the poloidal field $b^r b_r$ only becoming comparable in strength close to the event horizon, $r \lesssim 2 \, r_g$. Although the poloidal field is stronger than the toroidal field in the magnetized bubbles, we see that it plays a lesser role in the global structure of MADs as the averaged latitudinal magnetic field is quite weak compared to the toroidal and poloidal magnetic fields.

To better understand the steady-state structure of MADs, we express the general relativistic conservation of radial momentum in the following form:
\begin{multline}\label{eq:rad_eq}
    -\partial_\mu(\sqrt{-g}(\rho h +2p_\mathrm{mag})u^\mu u_r)+\partial_\mu(\sqrt{-g}b^\mu b_r) \\ \qquad +\sqrt{-g}\Bigl[-\partial_r(P+p_\mathrm{mag}) + \Gamma^\mu_{\mu_r}(\rho h +2p_\mathrm{mag})u^\mu u_\mu \Bigr. \\ 
     \Bigl. \qquad\qquad - \Gamma^\mu_{\mu_r}
     b^\mu b_\mu + \Gamma^t_{r r}T^r_t + \Gamma^r_{r t}T^t_r + \mathcal{O}\left(\frac{a}{M}\right)\Bigr] =0,
\end{multline}
where $\rho$ is the rest-mass density, $u^\mu$ the 4-velocity, $h$ the specific enthalpy, $P$ the thermal pressure, $p_\mathrm{mag}\equiv b^\mu b_\mu /2$ the magnetic pressure, $\Gamma^\nu_{\mu\lambda}$ the Christoffel symbol and $T^\mu_\nu$ the stress-energy tensor. At $r=20\:r_g$, where the relativistic effects are of second order, it is sufficient to identify the relevant physical terms of equation (\ref{eq:rad_eq}) in the Newtonian limit.  We associate the first term with the poloidal acceleration of the flow, the second term with the poloidal magnetic tension of the field lines, the third term with the gradient of the thermal and magnetic pressure, the spatial component of the fourth term with centrifugal-like effects, the temporal component of the fourth term with gravity-like effects, and the fifth term with the magnetic hoop stress. The remaining terms (dependent on the stress-energy tensor) vanish in the non-relativistic limit, far from the black hole. 

In the bottom panel of \autoref{fig:sim_forces} we plot the main contributions to the radial force budget, as well as the sum of all components in equation (\ref{eq:rad_eq}) as a black thick solid line. We see that the poloidal magnetic tension as well as the poloidal magnetic pressure are completely negligible at $20\:r_g$, with support against gravity coming primarily from thermal pressure and the centrifugal terms. We note that magnetic pressure (predominantly toroidal) does become the main term opposing gravity at $r<2\:r_g$, as can be seen on the top panel of \autoref{fig:sim_forces}. However, the disc is MAD to much larger radii than $2\:r_g$ and neither magnetic pressure nor magnetic tension is the dominant term in the radial equilibrium of the disc. These results are thus inconsistent with scenarios for MAD accretion that predict a magnetospheric radius where the poloidal magnetic tension or pressure balances gravity in the radial direction \citep{narayan2003,mckinney2012}. 

\subsection{Persistence of the magnetorotational instability in MADs}
\label{sec:MRI}

It is often stated that the strong poloidal field accumulated in MADs is able to quench MRI \citep{igumenshchev2008,mckinney2012,marshall2018,white2019}. This is based on the fact that, in MADs, the wavelength of the most unstable linear vertical mode is $\gtrsim H$, where $H$ is the height of the disc. We show here that this may not be sufficient to quench MRI and propose that MRI is still the main contributor to angular momentum transport in MADs.

\begin{figure}
\includegraphics[width=90mm]{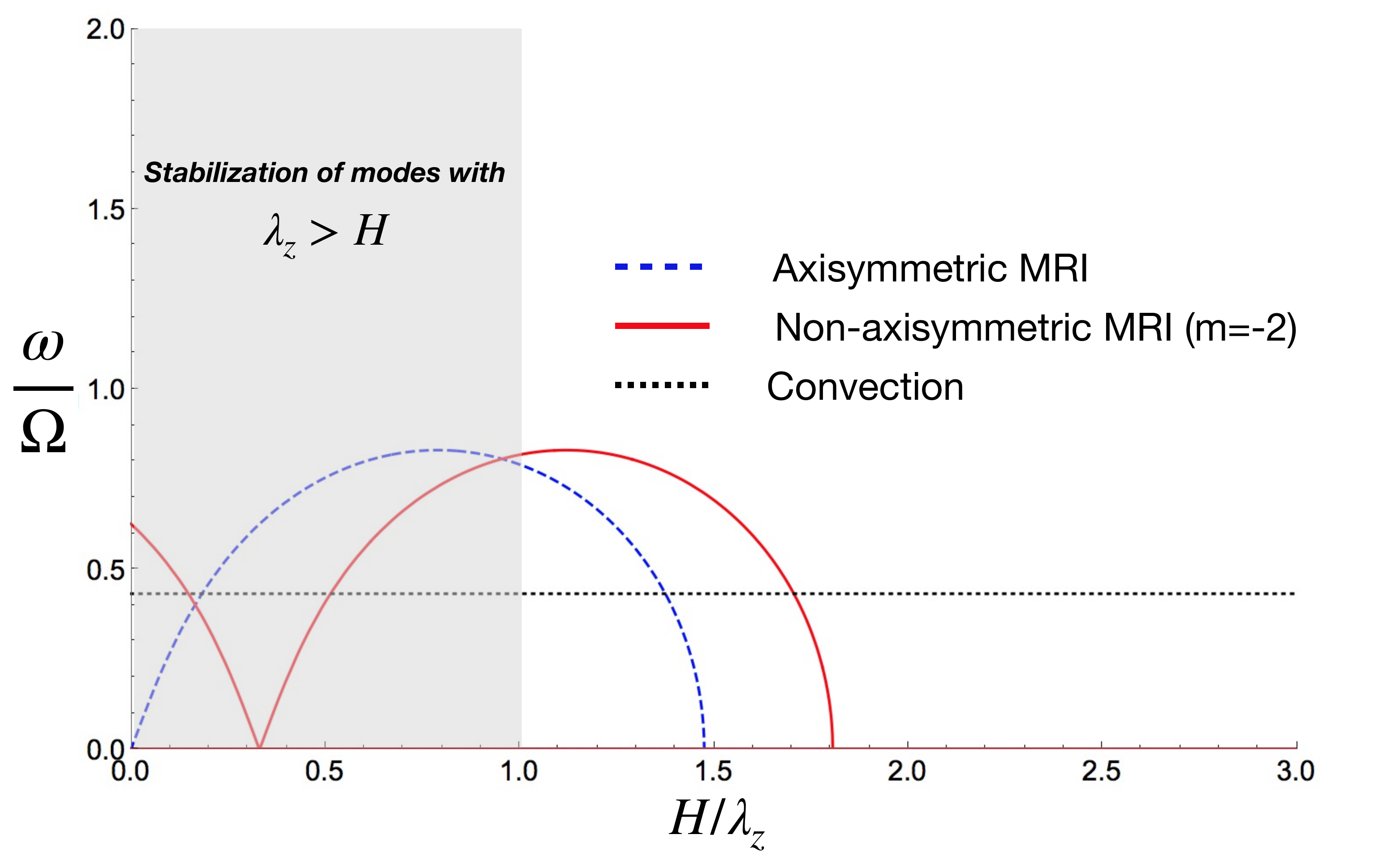}
\vspace*{-3mm}
\caption{Growth rate as a function of $H/\lambda_z$  for axisymmetric MRI (blue dashed line), non-axisymmetric MRI with the azimuthal wavenumber $m=-2$ (red solid line) and axisymmetric convection (black dotted line), assuming the background conditions measured on the midplane in our MAD simulation at $r = 20\:r_g$.  Although the most unstable axisymmetric linear mode of MRI is stabilized, there are slightly slower growing modes that are unstable in our MAD simulation. Plus, there always exist modes that are unstable to non-axisymmetric MRI. Note that convective modes can grow on timescales that are comparable to MRI, as we assume in \autoref{sec:Hoiland}.} 
\label{fig:dispersion_MRI} 
\end{figure}

\autoref{fig:dispersion_MRI} shows the growth rates of the unstable modes of axisymmetic MRI (dashed blue line), non-axisymmetric MRI with the azimuthal wavenumber $m=-2$ (solid red line) and radial convection (dotted black line) in the midplane as a function of $H/\lambda_z$, where $\lambda_z$ is the wavelength of a vertical mode. We compute the growth rates by solving equation [A3] of \cite{das2018} and by fixing all the free parameters in the dispersion relation like the strength of the vertical magnetic field, the strength of the azimuthal magnetic field, the gradient of the background entropy and the rotation rate using the values from our MAD simulation.\footnote{The calculation of the convective growth rate assumes that terms containing the azimuthal and vertical wavenumbers are negligible; see \autoref{sec:Hoiland}.} The grey shaded area denotes the zone where the vertical wavelength of a mode is larger than the scale height of the disc. It is usually assumed that in this limit MRI modes are stabilized \citep{balbus1991,balbus1998}. As shown in  \autoref{fig:dispersion_MRI}, for axisymmetric MRI, the most unstable mode is located in the shaded area as found previously \citep{mckinney2012,white2019}. However, MRI may still be alive in MADs for two reasons. First, even if the most unstable mode is in the shaded region there are still slightly slower growing modes that lie outside the shaded region. Second, even if all the modes of axisymmetric MRI were stabilized by a strong enough vertical magnetic field, one can easily find a maximally unstable, non-axisymmetric MRI mode with negative $m$ that lies outside the shaded region.  This is fundamentally due to the fact that there exists a degeneracy between modes growing on the vertical magnetic field and the azimuthal magnetic field in the dispersion relation of \cite{das2018}. This allows non-axisymmetric unstable MRI modes to exist for any $k_z$ as long as the toroidal field is weak enough.

\begin{figure}
\includegraphics[width=90mm]{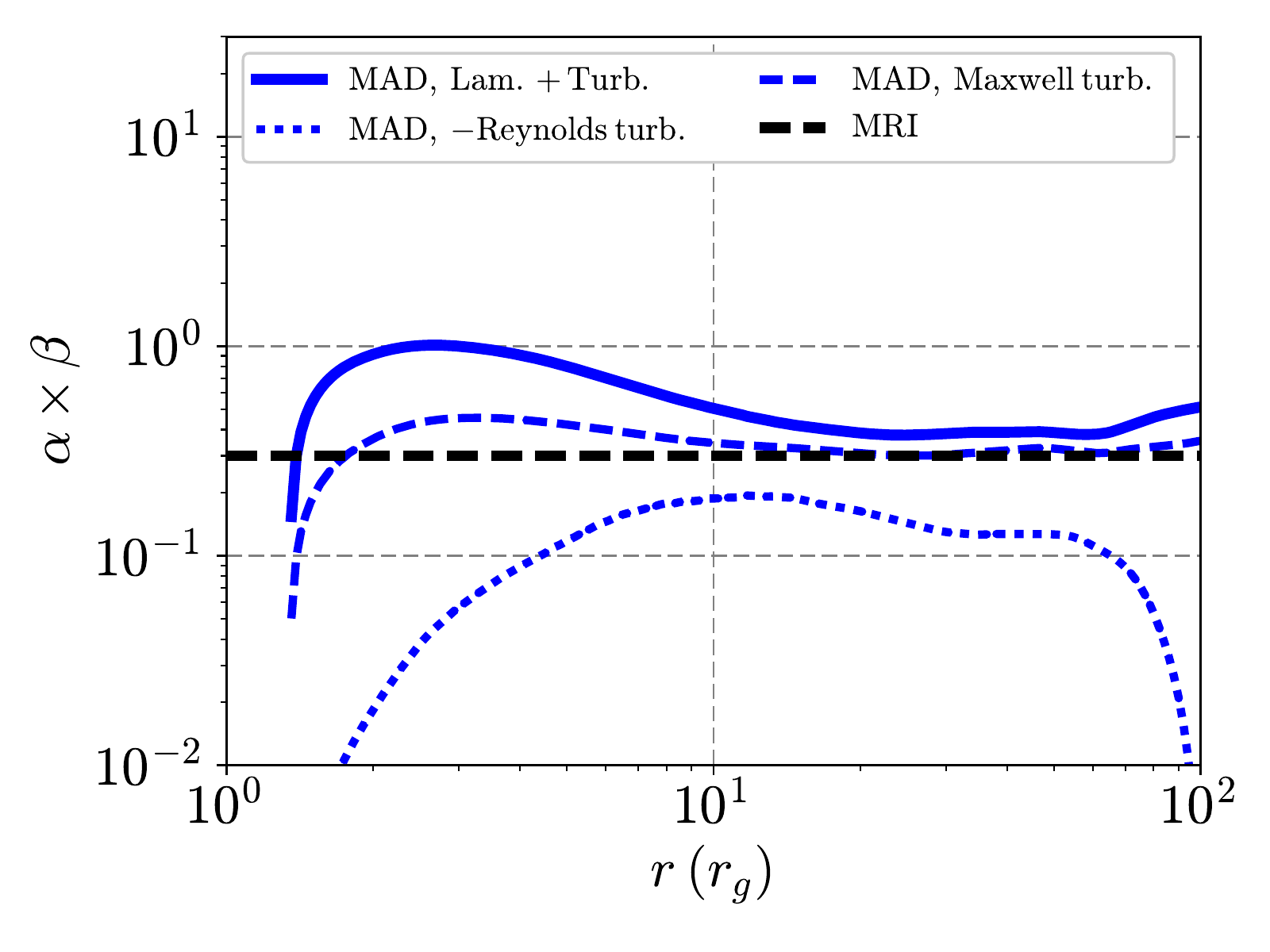}
\vspace*{-6mm}
\caption{Time- and spherically averaged $\alpha\times\beta$ measured in our MAD simulation (blue line) as a function of radius, where $\alpha$ includes the laminar and turbulent stresses. The dashed and dotted blue lines show the decomposition of the turbulent stress into the Maxwell and Reynolds stress, respectively. The dashed black line shows the prediction from MRI shearing-box simulations, $\alpha\times\beta\approx0.3$. Our measured values agree very well with the shearing-box prediction, especially for the component of $\alpha$ based on the turbulent Maxwell stress. This suggests that MRI is producing most of the angular momentum transport in our MAD simulation.}
\label{fig:alpha_simus}
\end{figure}

As further evidence, we plot in \autoref{fig:alpha_simus} the time- and spherically averaged values of the $\alpha$-parameter as defined in \cite{avara2016},
\begin{equation}
\alpha \equiv \langle \rho u_{\hat{r}} u_{\hat{\phi}} - b_{\hat{r}} b_{\hat{\phi}} \rangle_\rho/\langle P \rangle_\rho,
\end{equation}
times the plasma $\beta-$parameter, $\beta\equiv 2 \langle P \rangle_\rho / \langle b^\mu b_\mu \rangle_\rho$, measured in our MAD simulation as a function of radius. We also plot the decomposition of the turbulent stress into the Reynolds component and the Maxwell component defined, respectively, as 

\begin{equation}
\alpha_\mathrm{Rey,\:turb} \equiv (\langle \rho u_{\hat{r}} u_{\hat{\phi}} \rangle_\rho  -  \langle\rho\rangle \langle u_{\hat{r}} \rangle_\rho \langle u_{\hat{\phi}} \rangle_\rho)/\langle P \rangle_\rho,
\end{equation}

\begin{equation}
\alpha_\mathrm{Max,\:turb} \equiv ( - \langle b_{\hat{r}} b_{\hat{\phi}} \rangle_\rho  + \langle b_{\hat{r}} \rangle_\rho \langle b_{\hat{\phi}} \rangle_\rho)/\langle P \rangle_\rho.
\end{equation}

The dashed black line shows the empirical law $\alpha\times \beta\approx0.3$, derived from the results of shearing-box simulations of MRI \citep{hawley1995,salvessen2016}. We see that the turbulent Maxwell stress in the MAD simulation is in very good agreement with this empirical law while the turbulent Reynolds stress has a much smaller magnitude, also in agreement with MRI simulations.\footnote{We note, however, that the turbulent Reynolds stress is negative, which is not generally the case in MRI simulations. This might be due to convective angular momentum transport being important in the disc (\S\ref{sec:MADsim}).} The agreement between the empirical law and the MAD simulation gives additional support to the idea that MRI is still active in MADs.

\section{Convective stability of strongly magnetized discs}\label{sec:Hoiland}

We saw in \autoref{sec:sims} that the classical scenario of MADs --- in which sub-Keplerian rotation results from the large poloidal field strength and magnetic flux saturation at all radii is explained by the presence of magnetized, low-density bubbles --- is incompatible with simulations. In this section and the next we propose an alternative to that scenario, with an analytic model based on the convective instability of MADs. Tested against our MAD simulation, our model explains both the sub-Keplerian rotation and flux saturation in quantitative detail. 

For our model, we consider the convective stability of rotating, highly magnetized flows (``magnetotori''), generalizing the H\o iland criteria 
that are well-known from hydrodynamical studies of axisymmetric rotating stars and thick accretion discs \citep{goldreich1967,tassoul1978, begelman1982, blandford2004}.  This analysis assumes that a strong toroidal  field ($B_\phi$) has been created and maintained in the flow by the MRI which, as we saw in \autoref{sec:MRI}, remains a viable candidate for creating the toroidal field under MAD conditions, contrary to some claims in the literature.  

The fact that there is no dynamically attainable, marginally unstable state for MRI raises the question of whether it makes sense to consider states that are marginally unstable to other, concurrent instabilities \citep{balbus1998,balbus2001,hawley2001}.  We will conjecture that the convective instabilities discussed here are sufficiently fast, and operate on sufficiently different length scales from MRI, that they can indeed drive secular evolution of the flow structure toward the marginally unstable state. 

\subsection{Modified H\o iland criteria}\label{sec: Hoiland1}

In the usual derivation of the H\o iland stability criteria, a narrow but long ribbon of fluid, aligned in the azimuthal direction, is displaced in the poloidal plane by $d {\mathbf r}$, conserving both angular momentum and entropy. The ribbon is assumed to maintain pressure equilibrium with the local background flow, which has a distribution of specific angular momentum $L(r,\theta) \equiv \Omega R^2$ and entropy $S(r,\theta) \equiv \ln (P/\rho^\gamma)$, where $R= r\sin\theta$ is the cylindrical radius. One can then calculate the net centrifugal + bouyant acceleration on the ribbon in its displaced position and determine stability.  
The only difference between the situation in a magnetotorus and that in a hydrodynamical flow is that there is a third conserved quantity, 
\begin{equation}
    \label{eq:Kdef}
    K \equiv \ln \left( {B_\phi^2\over \rho^2 R^2}\right), 
\end{equation}
which follows from the flux-freezing condition.  It is straightforward to include conservation of $K$ in the stability calculation, and we closely follow the approach of \cite{begelman1982} in deriving the modifications below.       

For any fluid quantity $X$, denote the difference between the value of $X$ inside the displaced ribbon and  that in the ambient medium by
\begin{equation}
    \label{eq:deltaX}
    \delta X \equiv X_{\rm ribbon} - X_{\rm ambient} .
\end{equation}
For the conserved quantities $X_{\rm cons} = S, K, L$, we have 
\begin{equation}
    \label{eq:deltaXcons}
    \delta X_{\rm cons} = - \nabla X_{\rm cons} \cdot d{\mathbf r} ,
\end{equation}
and from the definitions of $S$ and $K$ we also have
\begin{equation}
    \label{eq:deltarho1}
    \delta S = {\delta P \over P}   - \gamma {\delta \rho \over \rho}; \ \ \delta K = 2{\delta B_\phi \over B_\phi} -  2{\delta \rho \over \rho}.
\end{equation}
Because the magnetotorus is supported by a combination of magnetic and gas pressure, the pressure balance condition in the hydrodynamical case, $\delta P =0$, is updated to 
 \begin{equation}
     \label{eq:Pcond}
     \delta\left(  P + {B_\phi^2\over 8\pi}  \right) = 0.
 \end{equation}
 
 From equations (\ref{eq:deltaXcons}), (\ref{eq:deltarho1}) and (\ref{eq:Pcond}), we can derive the density difference.  Using the plasma $\beta-$parameter $\beta \equiv 8\pi P/ B_\phi^2$, we have 
 \begin{equation}
     \label{eq:deltarho}
     {\delta\rho\over \rho} = {\left(\beta \nabla S + \nabla K \right)  \over  \gamma \beta + 2 } \cdot d {\mathbf r}.
 \end{equation}
The net acceleration due to buoyancy is then given by
\begin{equation}
    \label{eq:abuoy}
    \delta {\mathbf a}_{\rm buoy} = {\delta\rho\over\rho} {\mathbf g}_{\rm eff},
\end{equation}
where 
\begin{equation}
    \label{eq:geff}
    {\mathbf g}_{\rm eff} = {1\over\rho}\left[ \nabla P + {1\over R^2}\nabla \left({B_\phi^2 R^2\over 8\pi}\right)\right]
\end{equation}
is the effective gravity, while the centrifugal acceleration (unchanged from \citealt{begelman1982}) is  
\begin{equation}
    \label{eq:acent}
    \delta {\mathbf a}_{\rm cent} = -{1\over R^3} \left( \nabla L^2 \cdot \delta{\mathbf r}\right) \hat R.
\end{equation}
A sufficient condition for instability is then
\begin{equation}
    \label{eq:hoiland0}
    \left( \delta {\mathbf a}_{\rm buoy} + \delta {\mathbf a}_{\rm cent} \right) \cdot d{\mathbf r} > 0.
\end{equation}

As discussed in \cite{blandford2004}, equation (\ref{eq:hoiland0}) leads to two instability criteria. The first modified H\o iland criterion can be expressed as
\begin{equation}
    \label{eq:hoiland1}
     {\mathbf g}_{\rm eff} \cdot {\left(\beta \nabla S + \nabla K \right)  \over  \gamma \beta + 2 }  +  \nabla\left({1\over 2 R^2}\right) \cdot \nabla L^2 > 0, 
\end{equation}
while the second condition is
\begin{equation}
    \label{eq:hoiland2}
     \left[{\mathbf g}_{\rm eff} \times \nabla R \right] \cdot \left[\left(\beta\nabla S + \nabla K \right)  \times \nabla L \right] > 0. 
\end{equation}
Conditions (\ref{eq:hoiland1}) and (\ref{eq:hoiland2}) determine instability with respect to radial and tangential displacements, respectively; they reduce to the usual H\o iland instability criteria in the limit $B_\phi \rightarrow 0$.

\subsection{Marginally stable discs}

Our principal assumption is that dissipative and dynamo processes inside the flow, driven primarily by MRI over ``long'' timescales, also drive the system unstable under the modified H\o iland criteria, and that the flow is able to relax to a state close to marginal stability according to conditions  (\ref{eq:hoiland1}) and (\ref{eq:hoiland2}). 

The equilibrium structure of an axisymmetric magnetotorus in a Keplerian potential is governed by equation (\ref{eq:geff}) with 
\begin{equation}
    \label{eq:geff2}
    {\mathbf g}_{\rm eff} = \left(-{1\over r^2} + {L^2 \csc^2\theta \over r^3}, {L^2 \csc^2\theta \cot\theta\over r^3}\right)
\end{equation}
in spherical polar coordinates, adopting $GM=1$. 

An important difference between a magnetotorus and an unmagnetized hydrodynamical flow (e.g., \citealt{blandford2004}) is that the equilibrium and stability conditions depend on four fluid variables, e.g., $P, B_\phi, \rho, L$, instead of three (where we recall that $\beta$ depends on $P$ and $B_\phi$). Since there are two equilibrium equations, this leaves two additional constraints that need to be imposed (instead of one) in order to obtain a specific model. In this paper (and in the interest of mathematical simplicity), we will assume that one of these constraints is that $\beta = {\it constant}$. This turns out to be a good approximation for the purpose of comparing marginally unstable models to our MAD simulation (\autoref{sec:MADsim}).

Then, for systems that are marginally unstable according to condition (\ref{eq:hoiland2}), we will assume that
\begin{equation}
    \label{eq:mgt1}
    (\beta \nabla S + \nabla K ) \times \nabla L = 0 . 
\end{equation}
We term such systems {\it magnetogyrentropic}, by analogy with the gyrentropes discussed by \cite{blandford2004}. Alternatively, we term systems that are marginally unstable with respect to condition (\ref{eq:hoiland1}) {\it radially convective}. We will later argue that the latter models are associated with MADs.

\section{Self-similar models} 
\label{sec:selfsim}

Following \cite{blandford2004}, we adopt the self-similar scalings 
\begin{equation}
    \label{eq:selfsim}
    P = r^{n-{5\over 2}} p(\theta), \  {B_\phi^2\over 8\pi} = r^{n-{5\over 2}} p_B(\theta), \ \rho = r^{n-{3\over 2}} \bar\rho(\theta), \ L = r^{1\over 2} \ell(\theta),
\end{equation}
which allow us to write the equilibrium conditions as 
\begin{equation}
    \label{eq:selfsim2}
    \left({5\over 2} - n\right) {p\over \bar\rho} + \left({1\over 2} - n\right) {p_B\over \bar\rho}  = 1 - \ell^2 \csc^2\theta,  
\end{equation}
\begin{equation}
    \label{eq:selfsim3}
     {p'\over \bar\rho} + {p'_B\over \bar\rho} + 2 {p_B\over \bar\rho} \cot\theta  = \ell^2 \csc^2\theta \cot\theta,
     \end{equation}
where a prime denotes differentiation with respect to $\theta$.  Henceforth we will drop the bar over the angular function $\bar \rho (\theta) $. For $p = \beta p_B$ with constant $\beta$, these equations simplify to
\begin{equation}
    \label{eq:selfsim4}
   \left[ \left({5\over 2} - n\right)\beta  + \left({1\over 2} - n\right)\right] {p_B\over \rho}  = 1 - \ell^2 \csc^2\theta,  
\end{equation}
\begin{equation}
    \label{eq:selfsim5}
     (1 +\beta) {p'_B\over \rho} + 2 {p_B\over \rho} \cot\theta  = \ell^2 \csc^2\theta \cot\theta .   
\end{equation}

The parameter $n$ was adopted by \cite{blandford2004} to characterize self-similar, quasi-Keplerian discs with both inflow and outflow, with $\dot M \propto r^n$. Thus, $n=0$ corresponds to pure accretion and $n=1$ corresponds to a flow with $\dot M \propto r$ that liberates binding energy uniformly with radius.  In the absence of any other energy source we expect $0 \leq n < 1$, but note that a broader range could be accessible if, e.g., black hole spin energy is injected over a range of radii. We next consider the two types of marginally stable model in turn. 

\subsection{Magnetogyrentropic discs}\label{sec: mgt}
We first consider models for which Equation (\ref{eq:mgt1}) is satisfied.  These models are marginally unstable to the second H\o iland criterion, equation (\ref{eq:hoiland2}), and lie at the boundary for convection in the vertical direction.  These models turn out to be primarily of academic interest in the context of this paper, since neither the MAD nor the SANE model that we analyze lies close to this threshold.  However, hydrodynamical models satisfying this condition were discussed at length by \cite{blandford2004} and it is interesting to check the modifications introduced by the strong magnetic field.

Self-similarity demands that 
\begin{equation}
    \label{eq:mgt2}
   (\beta \nabla S + \nabla K) = - q {\nabla L \over L} ,
\end{equation}
where $q$ is a constant.  From the definitions of $S$ and $K$ and the radial scalings adopted above, one can readily verify that 
\begin{equation}
    \label{eq:mgt3}
   q =  \beta \left[(5-3\gamma) + 2 (\gamma-1)n\right] + (3 + 2 n ) .
\end{equation}
In the hydrodynamical limit, $\beta \rightarrow\infty$, equations (\ref{eq:mgt2}) and (\ref{eq:mgt3}) are consistent with equations [34] and [23] of \cite{blandford2004} (where we note that $S$ is defined somewhat differently).

The angular component of equation (\ref{eq:mgt2}) then yields one linear relationship among the logarithmic derivatives of $\rho$, $p_B$, and $\ell$, while differentiating equation (\ref{eq:selfsim4}) with respect to $\theta$ yields a second, allowing us to eliminate $\rho'/\rho$. Dividing equation (\ref{eq:selfsim5}) by equation (\ref{eq:selfsim4}) allows us to eliminate $p_B'/p_B$, yielding a first-order equation for $\ell(\theta)$.  This equation can be expressed compactly if we change the independent variable to $x\equiv \csc^2\theta$ and the dependent variable to $y(x) \equiv \ell^2 \csc^2\theta$. We also define the parameters
\begin{equation}
    \label{eq:mgt4}
    d \equiv {2 (2-\gamma)\beta\over 1+ \beta}; \ \ \ c\equiv 4 + 2\gamma\beta + q; \ \ \ w \equiv {q-d\over c}.
\end{equation}
After some algebra, we obtain
\begin{equation}
    \label{eq:mgt5}
    2 \left({q\over c} -  y\right) x {y'\over y} = 2 w- y,
\end{equation}
where a prime now denotes differentiation with respect to $x$.  We can solve equation (\ref{eq:mgt5}) implicitly for $x(y)$,   
\begin{equation}
    \label{eq:mgt6}
    x(y) = \left( {y \over \ell_0^2}\right)^{q\over q-d} \left( {y - 2 w \over  \ell_0^2 - 2 w }\right)^{q -2d\over q-d} ,
\end{equation}
where $\ell_0 = y(1)^{1/2}$ is the specific angular momentum on the midplane.  

These solutions can have a well-defined disc surface, where $p_B$, $p$ and $\rho$ vanish at $y = \ell^2 / \sin^2\theta = 1$, provided that both the numerator and denominator in the right-hand factor have the same sign.  The opening angle $\theta_d$ and surface angular momentum $\ell_d$ are given by
\begin{equation}
    \label{eq:mgt7}
    \sin\theta_d = \ell_d = \ell_0^{q\over q-d} \left( { \ell_0^2  - 2 w\over 1 - 2w}\right)^{q -2d\over 2( q-d)} .
\end{equation}
Equation (\ref{eq:mgt7}) generalizes equation [41] of \cite{blandford2004}, with $w$ replacing their parameter $a$.  For the special case $d/q \rightarrow 0$, which corresponds to both the purely hydrodynamical case considered by \cite{blandford2004} ($\beta \rightarrow \infty$) and the purely magnetic case $\beta = 0$, we can obtain an explicit solution,
\begin{equation}
    \label{eq:mgt8}
    \ell (\theta) = \left\{ w + \left[ w^2 + \ell_0^2 (\ell_0^2 - 2 w)\csc^2\theta \right]^{1/2}\right\}^{1/2} \sin\theta , 
\end{equation}
which is equivalent to equation [36] of \cite{blandford2004}. 

One can show that $w < 1/2$ for all $\gamma, \beta$ with $n< 1/2$ and for all except a small sliver of parameter space with $n>1/2$, mostly coincident with the regime $n > (5\beta + 1)/2$ in which the rotation on the midplane is super-Keplerian ($\ell_0 >1$) due to the effects of magnetic tension (cf.~equation \ref{eq:selfsim2}). For practical purposes, we will therefore assume that ``well-behaved" magnetogyrentropic models with a surface exist only for sufficiently high midplane angular momentum, $\ell_0 > (2 w)^{1/2} $. As $\ell_0$ approaches $(2 w)^{1/2}$ from above, the disc surface closes up to the rotation axis.

\begin{figure}
\includegraphics[width=85mm,height=85mm]{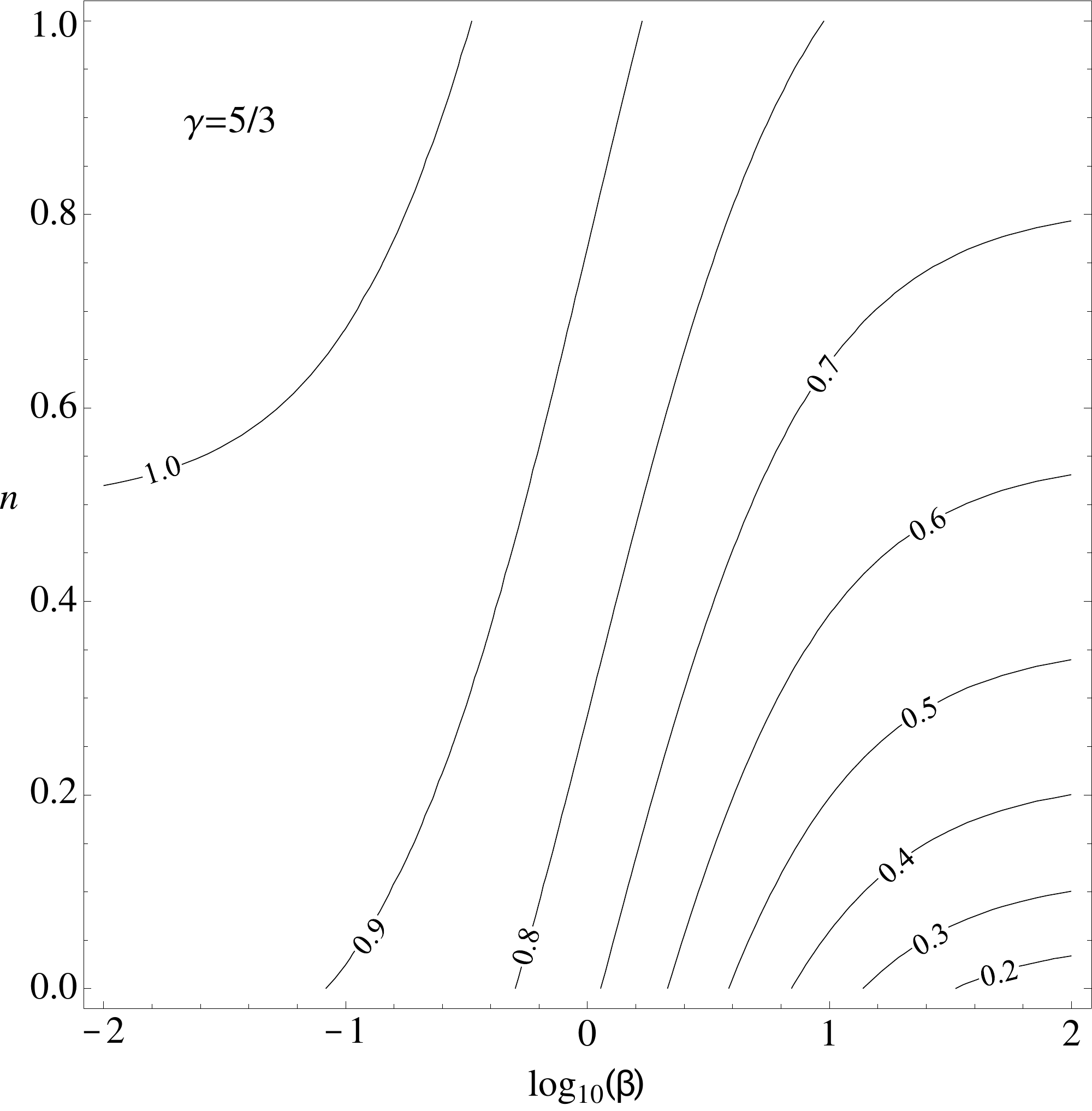}
\caption{Contours of minimum midplane angular momentum $\ell_0 = (2 w)^{1/2}$ consistent with a well-behaved magnetogyrentropic disc structure, for $\gamma = 5/3$. Magnetically dominated discs ($\beta < 1$) require near-Keplerian rotation because the magnetic tension compensates for the magnetic pressure gradient.}
\label{fig:ellzeromgt53}
\end{figure}

\begin{figure}
\includegraphics[width=85mm,height=85mm]{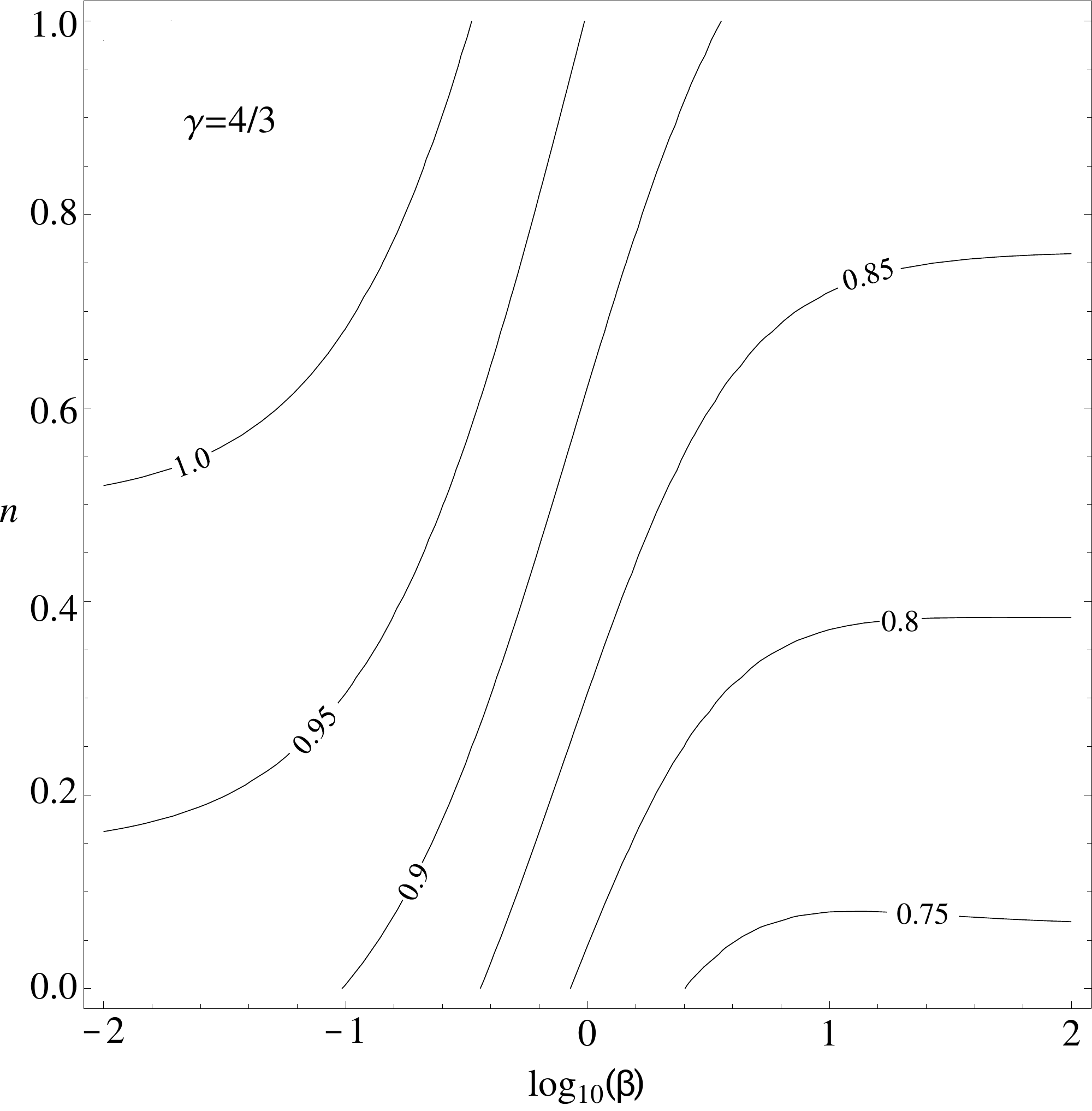}
\caption{Same as Figure \ref{fig:ellzeromgt53}, but for $\gamma = 4/3$. The softer equation of state means that these discs must be rapidly rotating even when dominated by gas pressure.}
\label{fig:ellzeromgt43}
\end{figure}

Figures \ref{fig:ellzeromgt53} and \ref{fig:ellzeromgt43} show contours of minimum $\ell_0$ for the magnetogyrentropic models with $\gamma = 5/3$ and 4/3, respectively, as a function of $\log \beta$ and $n$. For the magnetically dominated models ($\log \beta < 0$), the minimum values of $\ell_0$ are already very close to Keplerian, hence all such models must have  nearly Keplerian rotation speeds.  This is mainly due to the effects of magnetic tension, which partially (or even fully, for $n\geq 1/2$) cancels the radial magnetic pressure gradient in opposing gravity.  For the $\gamma = 4/3 $ models, the discs are rapidly rotating even when gas pressure is dominant because of the tendency of softer equations of state to lead to stronger convective instability. 
 
\subsection{Radially convective discs}\label{sec:radconv}
\begin{figure}
\includegraphics[width=87mm,height=87mm]{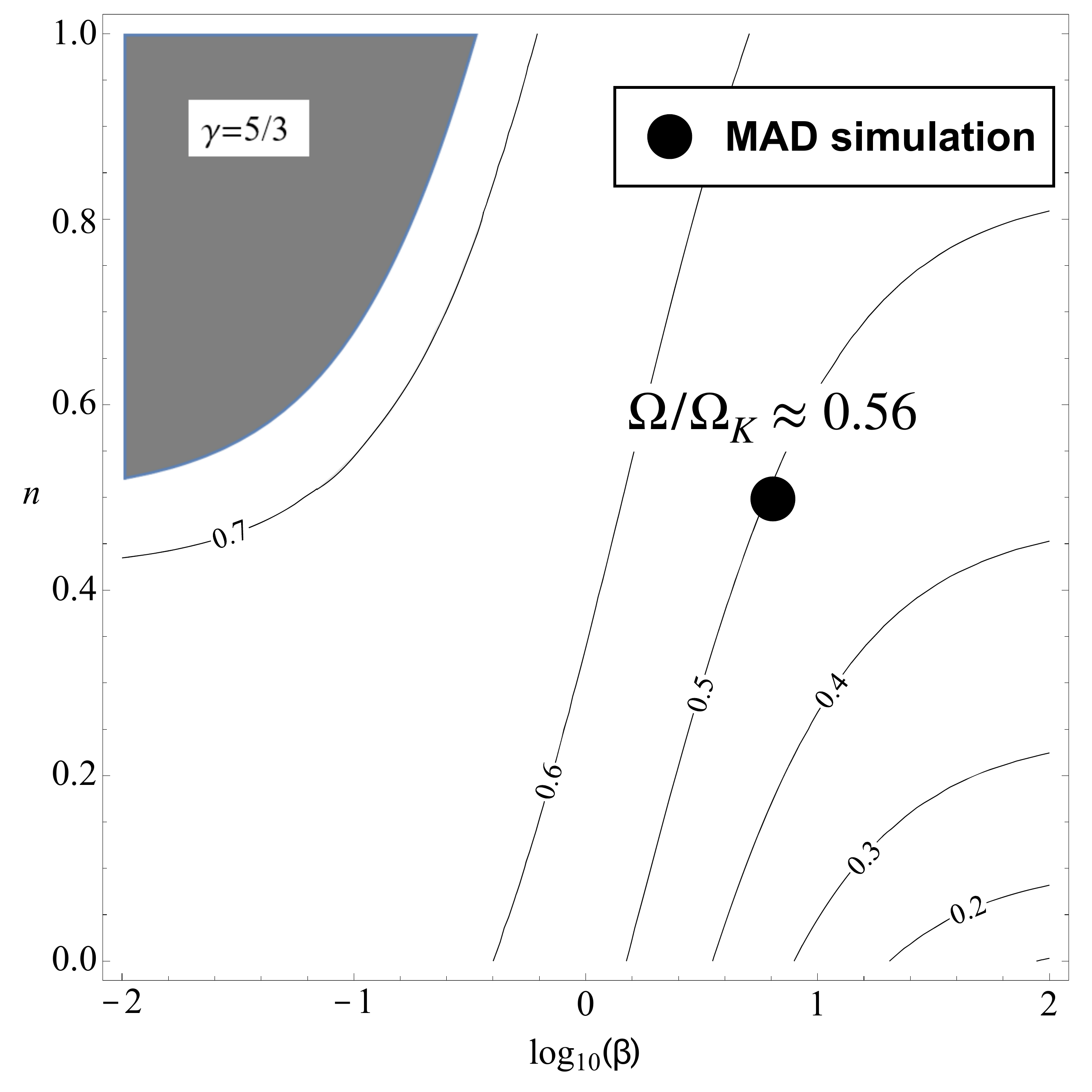}
\caption{Contours of $\ell_0 = (q/c)^{1/2}$ for discs marginally unstable to radial convection with $\gamma = 5/3$. The midplane angular momentum is lower than the minimum angular momentum for a magnetogyrentropic disc by a factor $\approx \sqrt{2}$ in both the high- and low-$\beta$ limits.  The shaded region corresponds to $w > 1/2$, where no well-behaved solutions exist. The $\gamma = 4/3$ case is qualitatively similar, but with all the angular momenta shifted to higher values. The large dot, located at values of $n$ and $\beta$ measured from our MAD simulation, lies very close to the $\ell_0 = 0.5$ contour; the measured value for the simulation is $\ell_0 = \Omega/\Omega_{\rm K} \approx 0.56$. }
\label{fig:ellmidrad53}
\end{figure}

We next consider discs that marginally satisfy condition (\ref{eq:hoiland1}). In terms of the variables and parameters adopted in the previous section, the structural equation can be written as 
\begin{equation}
    \label{eq:radconv1}
    2\left(1-{q\over c}\right)(x-1) x y' + (x-1)y [y - 2(1-w)] + \left({q\over c} - y\right)(1-y) = 0. 
\end{equation}
The first two terms vanish on the midplane ($x=1$), implying that the midplane angular momentum has the unique value $\ell_0 = (q/c)^{1/2}$, which equals $w^{1/2}$ in both the strongly and weakly magnetized limits. Thus, there is an angular momentum {\it gap} between radially convective discs, which have $\ell_0 \approx w^{1/2}$, and magnetogyrentropic discs with $\ell_0 > (2 w)^{1/2}$.  \autoref{fig:ellmidrad53} shows contours of midplane angular momentum for radially convective discs with $\gamma = 5/3$.

For $w < 1/2$, $y < 1$, one can show that $y$ is a monotonically increasing function of $x$.  Therefore, one can determine the existence of a disc surface by considering the limit $x \rightarrow \infty$, which has an attractor with $y'\rightarrow 0$, $y \rightarrow y_\infty = 2 (1-w)$.  Since $y_\infty > 1$ for $w < 1/2$, we see that a disc surface at $y=1$ must exist for all $w < 1/2$, i.e., virtually the entire parameter space of interest. Conversely, there are no solutions for $w > 1/2$.   
 
\begin{figure}
\includegraphics[width=85mm,height=58mm]{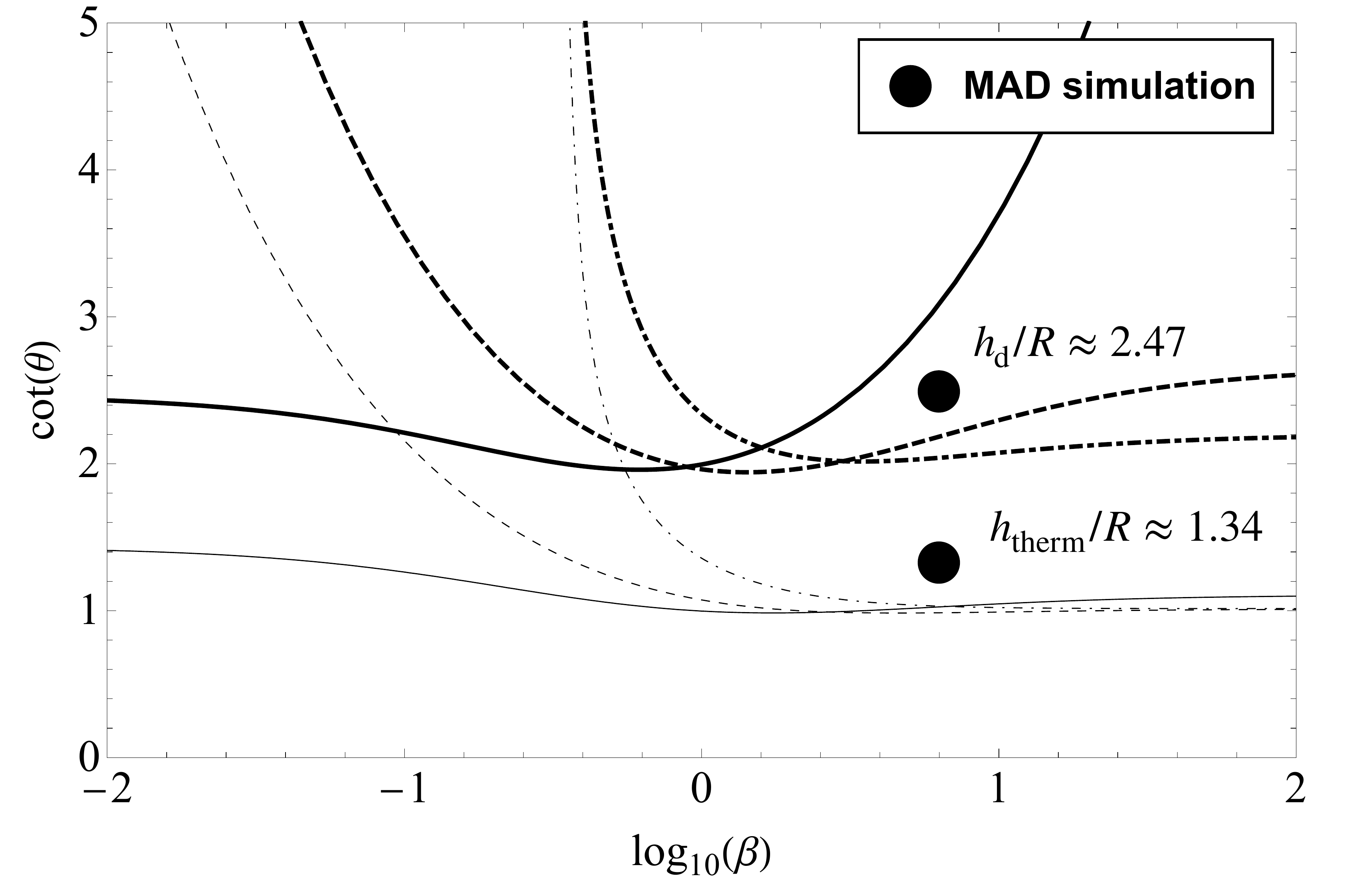}
\caption{Geometric thickness of radially convective disc models with $\gamma=5/3$ and $n=$ 0 (solid), 1/2 (dashed), and 1 (dot-dashed) as a function of $\beta$. The vertical axis shows the cylindrical aspect ratio $z/R = \cot\theta$. Thick curves show the total disc thickness $\theta_d$ while thin curves show the dimensionless pressure scale height measured along cylinders, i.e., the value of $z/R$ at which the pressure is $e^{-1}$ times that on the midplane. Large dots indicate values of the total thickness and thermal scale height (which closely approximates the pressure scale height), measured from our MAD simulation (cf.~\autoref{sec:MADsim}).}
\label{fig:radconv1v4}
\end{figure}

Radially convective disc solutions are geometrically quite thick for all parameters, as shown in \autoref{fig:radconv1v4}.  The discs are thinnest for $\beta$ of a few, with miminum total disc thickness (where pressure and density vanish) $z_d/R = \cot \theta_d \approx 1.8$, corresponding to an opening angle of $\approx 60$ degrees with respect to the midplane (i.e., $\theta_d \approx 30^\circ$). Disc thickness is particularly sensitive to $n$ in the magnetically dominated limit, with the largest $n$ yielding the thickest discs due to the effects of magnetic tension which, as noted before, tends to cancel the radial pressure gradient without affecting vertical support. 

 Although equation (\ref{eq:radconv1}) is easily integrated numerically, it is possible to construct a simple analytic formula that captures the behavior of $y(x)$ for all parameters of interest and  values of $x$, typically to better than 10 percent. Defining for convenience 
 \begin{equation}
     \label{eq:w2def}
     w_2 \equiv {q\over c}; \ \ \ f\equiv {w_2 [2(1-w) - w_2]\over (1-w_2)  },
 \end{equation}
the expression 
 \begin{equation}
     \label{eq:radconvapp}
 y = w_2 + {(2-2w-2w_2) f (x-1) \over 2-2w-2w_2 + f(x-1)}    
 \end{equation}
 interpolates accurately between the behavior near the midplane and the asymptotic limit at large $x$. It also yields an accurate analytic estimate of the disc thickness, 
  \begin{equation}
     \label{eq:radconvapp2}
 {z_d\over R} = \cot\theta_d =   { (1-w_2) \over w_2^{1/2} (1-2w)^{1/2}} .
 \end{equation}
 For all cases, $w\approx w_2$ to within about 10\%; thus, substituting $w_2$ for $w$ in all the expressions above also yields acceptable fits.   
 
 Since it allows analytic integration of the momentum equations, equation (\ref{eq:radconvapp}) is also useful for locating isobars and isodensity surfaces inside the disc.  Because $y$ varies slowly with $x$ near the disc surface, the pressure and density scale heights (defined as the place where the pressure or density drops to a factor $e^{-1}$ below the midplane value) can lie at much smaller latitudes than the disc surface.  The thin lines in  \autoref{fig:radconv1v4} show the pressure scale height calculated using the analytic fit; the density scale height lies between the pressure scale height and $x_d$.  For $\beta \gtrsim O(1)$, the pressure scale height surface lies about $45^\circ$ above the midplane. Interestingly, we note that the pressure in the magnetically dominated solutions can sometimes be non-monotonic, increasing weakly with height in the vicinity of the midplane while decreasing as expected at larger $x$.  We attribute this curious behavior to the variable effects of magnetic tension with height.

\subsection{Significance of the Bernoulli function}\label{sec:Bernoulli}
  
\cite{blandford2004} noted that their models of self-similar gyrentropic discs close up to the rotational axis exactly when the Bernoulli function on the midplane, $B_0$, vanishes, and that no well-behaved solutions exist for $B_0 > 0$. The Bernoulli function for a rotating, axisymmetric flow with negligible poloidal velocity is usually written 
\begin{equation}
    \label{Bernoullidef}
    B = H + {L^2\over 2 R^2} + \phi,
\end{equation}
where $H$ is the specific enthalpy and $\phi$ is the gravitational potential.  For an ideal gas with adiabatic index $\gamma$, $H = \gamma P / (\gamma - 1) \rho $.  At the free boundary of a gyrentrope $H = 0$, and the condition of dynamical equilibrium  then implies $B = B_d = - (2 r)^{-1} < 0$ on the disc surface. But one can also show that every ``isobern'' of fixed $B$ in a gyrentrope that reaches the disc surface must also pass through the midplane, implying that $B_0$ must be negative in order for the gyrentrope to have a surface.  Physically, the  combination of a positive Bernoulli function and a free boundary would enable the disc to disperse on a dynamical timescale.

Since the magnetogyrentropes we study in \autoref{sec: mgt} have similar generic properties to gasdynamical gyrentropes, we suspect that an analogous condition must hold.  However, our magnetized structures are subject to two additional complications: first, the directionality of the Lorentz force does not permit the derivation of a simple Bernoulli equation showing that the Bernoulli function is manifestly conserved in an adiabatic potential flow, and second, our assumption of a constant $\beta$ constrains the transfer of energy between gas and magnetic reservoirs along a streamline.     

To determine an appropriate form for the Bernoulli function in a magnetogyrentrope, we work backwards under the assumption that $B_0$ vanishes when $\ell_0^2 = 2 w $.  Setting $H r \equiv a p_B/\rho$ in self-similar variables, and using the radial momentum equation to express $(p_B/\rho)_0$ in terms of $\ell_0^2$, we find that
\begin{equation}
    \label{eq:Bernoullia}
    a = {2 + 4\beta + \beta^2\gamma \over 1 + \beta(\gamma - 1)},
\end{equation}
which reduces to the gasdynamical limit when $\beta \rightarrow \infty$ and yields $H = 2 P_B/\rho$ when gas pressure is negligible.  The latter is expected for a magnetic field since the pressure equals the magnetic energy density $U_B$, giving the familiar result $H= (P_B + U_B)/\rho$.  

For a radially convective disc $B_0 > 0$, since $\ell_0 < 2 w$, but we have seen that these flows nevertheless have a well-defined disc surface, with $B_d < 0$.  This is possible only if there is an internal surface at fixed $\theta$ with $B=0$, i.e., the isoberns do not not connect the disc surface to the midplane in radially convective flows. 
We can gain more insight by using equation (\ref{eq:Bernoullia}) and the radial momentum equation to write the Bernoulli function in the form
\begin{equation}
    \label{eq:Bernoulli1}
    B  = {2 \over (1 - 2 w) r } (2 w -y) .
\end{equation}
The gradient of $B$ is then
\begin{equation}
    \label{eq:Bernoulligrad}
      \nabla B  = {2 \over (1 - 2 w) r^2 } \left[y - 2 w, \  2 \cot\theta x y' \right].
\end{equation}
For well-behaved magnetogyrentropes we have $y - 2w > 0$ at all points, confirming that $\nabla B \times \hat \theta$ never vanishes and that the surfaces of constant $B$ connect the midplane to the disc surface.  For radially convective discs, however, $(\nabla B)_r$ is negative on the midplane but positive on the surface, implying that the constant-$B$ surfaces are aligned with the radial direction at some $\theta$ inside the disc.  Thus, there is no connection between the regions of positive $B$ near the midplane and the negative-$B$ layers near the surface.  These discs are therefore well-behaved despite having a positive Bernoulli function near the midplane.  

There is one interesting difference between our magnetogyrentropic models and the gyrentropes studied by \cite{blandford2004}. Using the angular momentum gradient 
\begin{equation}
    \label{eq:Lgrad}
      \nabla L^2  = {y \over x } \left[1, \  2 \cot\theta \left( 1-x {y'\over y} \right)\right] ,
\end{equation}
we can calculate 
\begin{equation}
    \label{eq:Bernoulligrad2}
     \left( \nabla B \times \nabla L^2\right)_\phi \propto 2x{y' \over y } (w-y) + y-2w  = - 2 {d\over c} x{y'\over y},
\end{equation}
where the last relation is obtained using equation (\ref{eq:mgt5}).  Thus, in contrast to a gyrentrope, where the $B$, $S$ and $L$ surfaces always coincide, for a magnetogyrentrope this is true only in the limit $d \rightarrow 0$, i.e., in the limits where either gas pressure or magnetic pressure dominates, but not in between.  
  
\subsection{Are MADs convectively unstable?}\label{sec:MADsim}

We can test our hypothesis that MADs are convectively unstable by evaluating the modified H\o iland criteria for the MAD simulation discussed in \autoref{sec:sims}. For comparison, we evaluate the same criteria for our SANE simulation as well.  

\begin{figure}
\includegraphics[width=90mm]{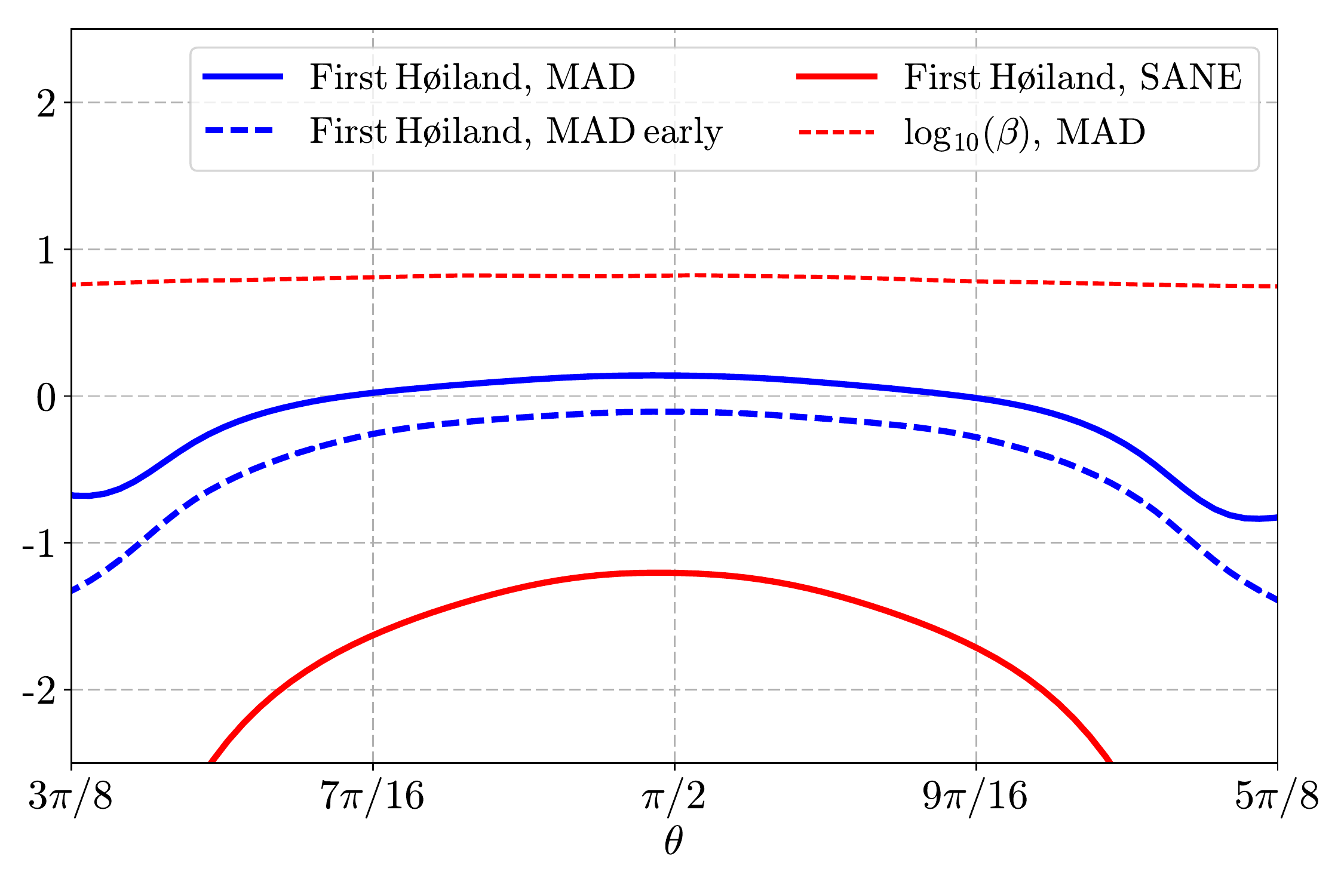}
\caption{First H\o iland criterion as a function of $\theta$ for the MAD (blue line) and SANE (red solid line) simulations at $r=30\:r_g$. The dashed blue line shows the first H\o iland criterion in our MAD simulation at an earlier time between $20,000$ and $30,000$ $r_g/c$.} The quantity plotted is the left-hand side of equation (\ref{eq:hoiland1}) divided by its first term, which represents the contribution of buoyancy. The small but positive values for the MAD simulation indicate convective instability to radial motions, close to marginal instability. In contrast, the SANE simulation is stable. Note that the MAD simulation is not convectively unstable at earlier times. The dashed red line shows the value of $\log_{10}(\beta)$ as a function of $\theta$ from our MAD simulation, validating  the assumption of contant $\beta$ in our analytic models.
\label{fig:Hoiland_MAD}
\end{figure}

In \autoref{fig:Hoiland_MAD} we plot the left-hand side of the first H\o iland criterion, equation (\ref{eq:hoiland1}), from the results of our MAD and SANE simulations, as a function of $\theta$. These quantities are azimuthally and time-averaged, and evaluated at a radius of $30\:r_g$. For the MAD case, we see that the first H\o iland criterion is positive around the midplane for $7\pi/16<\theta<9\pi/16$, indicating radial convective instability. In contrast, in the SANE simulation the disc is stable to radial convective motions. \cite{narayan2012} similarly analyzed the convective stability of MAD and SANE simulations but found less clearcut evidence for instability near the midplane over a wide range of radii in their MAD.  This is possibly due to their neglect of magnetic effects in the H\o iland criterion; these effects tend to increase instability at a given rotation rate.  

Note that our MAD simulation was not convectively unstable at $r=30\:r_g$ between $20,000$ and $30,000$ $r_g/c$, as shown by the dashed blue line in \autoref{fig:Hoiland_MAD}. This is consistent with \autoref{fig:flux_steady} and \autoref{fig:rotation}, where we showed that the MAD state gradually builds up. At these earlier times the MAD is rotating faster, the flux has not built up to its saturated value and the disc is not yet convectively unstable. We have verified that both simulations are stable close to the midplane with respect to the second H\o iland criterion, equation (\ref{eq:hoiland2}).  

While the stability results plotted in  \autoref{fig:Hoiland_MAD} do not assume self-similarity, we find that the structure of our simulated MAD closely approximates a marginally stable, self-similar model in several respects. To make this comparison, we first must determine the best fit to the parameter $n$, defined in equation (\ref{eq:selfsim}), and the plasma-$\beta$ parameter, which is assumed to be constant.

From the top panel of \autoref{fig:sim_forces}, we can constrain the radial slopes of the various MHD quantities and so the value of $n$. We see that a value of $n=1/2$ gives a very good fit to our results, at least for $r>10\:r_g$. For $r<10\:r_g$, the gas pressure and density still follow a power-law with index corresponding to $n = 1/2$ but the magnetic components deviate from this trend. 

It is interesting to note that $n=1/2$ is a special value where the toroidal magnetic pressure gradient and the hoop stress cancel each other. This is also what we find with our radial force budget in \autoref{sec:sims_measurements}. This means that overall the toroidal field does not exert a radial force on the disc and is consistent with the idea that it is able to approach a radial force-free configuration through turbulent diffusion.  

In \autoref{fig:Hoiland_MAD} we also plot the latitudinal profile of $\mathrm{log_{10}}(\beta)$. We see that our assumption of a constant $\beta$ in the disc is very well justified with $\mathrm{log_{10}}(\beta)\approx0.8$ {in the MAD simulation} for $5\pi/16<\theta<11\pi/16$. Plotting our best-fit values on the $n-\beta$ parameter plane of \autoref{fig:ellmidrad53}, we see that the measured value of the angular velocity from our MAD simulation, $\Omega \approx 0.56$ in Keplerian units, is comparable to the value predicted for marginal instability, $\Omega \approx 0.5$. The angular velocity was radially averaged from \autoref{fig:rotation} between the innermost circular orbit and $70\:r_g$. We also plot on \autoref{fig:radconv1v4} the geometrical aspect ratio,
\begin{equation}
\frac{h_d}{R} = \langle |\theta-\theta_0|\rangle _\rho,
\end{equation}
where $\theta_0$ is defined as 
\begin{equation}
\theta_0 = \frac{\pi}{2} + \langle \theta-\pi/2\rangle _\rho,
\end{equation}
and the thermal aspect ratio of the simulated MAD,
\begin{equation}
    \frac{h_\mathrm{th}}{R}\equiv \cot(\theta_\mathrm{th}-\pi/2),
\end{equation}
where $\theta_\mathrm{th}$ is the angle at which the thermal pressure has decreased by $1/e$ from the midplane. We see that our MAD simulation matches well the analytic prediction for the opening angle and pressure scale height in a marginally unstable disc. Taken together, the agreement between the analytic theory and our MAD simulation favors radial convective motions as the mechanism behind magnetic flux saturation in MADs.

\section{Discussion and Conclusions}\label{sec:discussion}

We have argued that the saturation of net magnetic flux in a hot, magnetically arrested disc is driven not by interchange instabilities associated with the poloidal field, but rather by radial convective/interchange instabilities triggered by a combination of gas entropy gradients and a dominant toroidal field.  This mechanism still requires a large enough poloidal field in order to operate, but only indirectly in the sense that the poloidal field is necessary to stimulate the growth and maintenance of a strong toroidal field through dynamo action.  Contrary to claims in the literature, we find that MRI is not suppressed in MADs, especially since nonaxisymmetric modes can grow rapidly once the a toroidal field is present, and we suggest that MRI is the most likely mechanism for maintaining the dominant $B_\phi$. 

To test the plausibility of this proposal, we generalized the H\o iland criteria to include the effects of a dynamically significant toroidal field and showed that a long-duration, large-dynamic-range MAD simulation is unstable to radial convection according to the revised criteria.  By comparison, a SANE simulation, i.e., without saturated flux, is stable to both radial and vertical convection.  

We also used the generalized criteria to derive radially self-similar models for marginally stable discs.  Remarkably, our MAD simulation closely matches several features of these self-similar models, including the rotation rate, pressure scale height, and total disc thickness. This suggests that convection may be driving the disc structure toward marginal instability.  Although the marginal state depends on $B_\phi$ and not directly on the net poloidal flux, the two are linked through the dynamo process needed to maintain the toroidal field.  

Perhaps the most striking characteristic of a MAD is the significantly sub-Keplerian midplane angular velocity of $\approx 0.5-0.6$ in Keplerian units, in contrast to SANE and other strongly magnetized discs \citep[e.g.,][]{mishra2020}, which invariably have rotation rates very close to Keplerian.  We suggest that the appearance of such a sub-Keplerian rotation rate may be a robust indicator that a given region of an accretion disc has reached flux saturation, i.e., has attained a MAD state.  The observed radial scaling of the toroidal field, $B_\phi \propto r^{-1}$, may also be a signature of strong radial diffusion of the field, since it corresponds to a balance between the radial magnetic pressure gradient and hoop stress.

Our argument depends on the speculative assertion that convective/interchange  instabilities operate fast enough to largely shape the disc structure, despite our claim that MRI cannot be turned off and operates concurrently.  Thus, technically there is no marginally unstable state and we must rely on convective instabilities leading to more effective flux diffusion than MRI.  This needs to be checked, although we note that the growth rates of convective instability and MRI are similar (\autoref{fig:dispersion_MRI}) and the values of $\alpha$ we derive from our MAD simulations are modest ($\sim O(0.1)$: cf.~ \autoref{fig:alpha_simus} and \autoref{fig:Hoiland_MAD}). Moreover, a geometrically thick disc subject to MRI-driven turbulence without convection is not expected to regulate its flux \citep{lubow1994}. Thus, we suggest that the secular evolution of disc structure could be regulated by convective instability,  even if MRI is responsible for regulating the dynamo and transporting angular momentum.  
 
Our analytic model applies to the steady-state structure of MADs at large distances from the BH ($r>10\:r_g$) and aims at explaining the characteristics of MADs, such as the saturation of flux and the sub-Keplerian rotation, at these large radii. Close to the black hole our Newtonian analytical model is not valid anymore and the self-similar trends followed by the gas density, gas pressure and magnetic pressure seem to deviate from the ones at larger radii. Moreover, very close to the black hole, at $r<2\:r_g$, the magnetic pressure becomes the dominant support of the accretion flow against gravity. This could cause the flow to be unstable to the Rayleigh-Taylor instability as claimed in the literature \citep{narayan2003,mckinney2012}. We leave this to further work but emphasize that even if the Rayleigh-Taylor instability does play a role close to the black hole it is unlikely to explain the behavior of MADs at larger radii.

Our analysis applies to ``hot'' MADs in the sense that the gas pressure is at least comparable to the pressure of the toroidal magnetic field (in our simulation it is several times larger). Under certain conditions it is likely that a ``hot'' MAD will cool rapidly \citep{dexter2021}, losing thermal pressure while retaining a large field.  It is doubtful that such a flow would remain convectively unstable near the midplane, given that the rotation rate would probably revert quickly to a near-Keplerian value.  The question of whether MADs would persist under these conditions, and what would determine flux saturation, remains to be explored. 
      
\section*{Acknowledgements}
We thank the referee for valuable comments and suggestions that improved the paper. We acknowledge financial support from  NASA Astrophysics Theory Program grants NNX16AI40G, NNX17AK55G, and 80NSSC20K0527 and NSF Grant AST-1903335, and by an Alfred P.~Sloan Research Fellowship (JD). The calculations presented here were carried out using resources supported by the NASA High-End Computing (HEC) Program through the NASA Advanced Supercomputing (NAS) Division at Ames Research Center. 

\section*{Data Availability}
The axisymmetrized data analyzed in this article will be shared on reasonable request to the corresponding author.




\bibliographystyle{mnras}
\bibliography{biblio} 


\bsp	
\label{lastpage}
\end{document}